\documentclass[10pt,twocolumn,oneside,final]{IEEEtran}

\usepackage{cite}
\usepackage{graphicx}
\usepackage{amsmath}
\usepackage{times}
\usepackage{latexsym}
\usepackage{bm}
\usepackage{amssymb}
\usepackage[center]{caption2}
\usepackage{array}
\usepackage{fancyhdr}
\usepackage{cite,graphicx,amsmath,amssymb}
\usepackage{citesort}
\usepackage{psfrag}
\usepackage{multirow}

\ifCLASSINFOpdf

\else

\fi

\newtheorem{corollary}{Corollary}
\newtheorem{prop}{Proposition}
\newtheorem{alg}{Algorithm}

\newcommand{\tr}{\text{tr}}

\makeatother

\begin{document}
\title{Linear Precoder Design for MIMO Interference Channels with Finite-Alphabet Signaling}

\author{\IEEEauthorblockN{Yongpeng Wu,~\IEEEmembership{Student Member,~IEEE,}
Chengshan Xiao,~\IEEEmembership{Fellow,~IEEE,} Xiqi Gao,~\IEEEmembership{Senior Member,~IEEE}, \\ John D. Matyjas,~\IEEEmembership{Member,~IEEE},
 and Zhi Ding,~\IEEEmembership{Fellow,~IEEE}}

\thanks{Manuscript received Feb. 15, 2013; revised May 26, 2013.
The associate editor coordinating the review of this paper and approving it
for publication was Tony Q. S. Quek.}

\thanks{Part of the material in this paper will be presented at IEEE GLOBECOM
2013. The work of Y. Wu and X. Gao was supported in part by
National Natural Science Foundation of China under Grants 60925004 and 61222102,
the China High-Tech 863 Plan under Grant 2012AA01A506,
and National Science and Technology Major Project of China under Grants 2011ZX03003-001 and 2011ZX03003-003-03.
The work of C. Xiao was supported in part by National Science
Foundation under Grants CCF-0915846 and ECCS-1231848.
The work of Z. Ding was supported by NSF Grants 0917251 and 1147930.
This work began while Y. Wu was a
visiting Missouri University of Science and Technology and was completed while
C. Xiao was a summer faculty fellow at the Air Force Research Laboratory in Rome, NY.
Approved for Public Release; Distribution Unlimited: 88ABW-2013-1062.
}
\thanks{ACKNOWLEDGMENT OF SUPPORT AND DISCLAIMER:
(a) Contractor acknowledges Government's support in the publication of
this paper. This material is based upon work funded by the AF Summer
Faculty Fellow Program, under AFRL Contract No. FA8750-11-2-0218.
(b) Any opinions, findings and conclusions or recommendations expressed
in this material are those of the author(s) and do not necessarily
reflect the views of AFRL.}

\thanks{Y. Wu and X. Gao are with the National Mobile Communications Research Laboratory,
Southeast University, Nanjing, 210096, P. R. China (Email:
ypwu@seu.edu.cn; xqgao@seu.edu.cn). }

\thanks{C. Xiao is with the Department of Electrical and Computer Engineering,
Missouri University of Science and Technology, Rolla, MO 65409, USA (Email: xiaoc@mst.edu). }

\thanks{J. D. Matyjas is with the Air Force Research Laboratory/RIT, Rome, NY
13441, USA (E-mail: John.Matyjas@rl.af.mil).}

\thanks{Z. Ding is with the Department of Electrical and Computer Engineering,
University of California, Davis, CA 95616, USA (Email: zding@ucdavis.ca).}

}

\maketitle

\begin{abstract}
This paper investigates the linear precoder design for
$K$-user interference channels of multiple-input
multiple-output (MIMO) transceivers under finite alphabet
inputs. We first obtain general explicit expressions of the
achievable rate for users in the MIMO interference channel systems.
We study optimal transmission strategies in both low and
high signal-to-noise ratio (SNR) regions. Given finite
alphabet inputs, we show that a simple power allocation design achieves optimal
performance at high SNR whereas the well-known interference alignment technique for Gaussian inputs
only utilizes a partial interference-free signal space for transmission and leads
to a constant rate loss when applied naively to finite-alphabet inputs.
Moreover,  we establish
necessary conditions for the linear precoder design to achieve weighted
sum-rate maximization. We also present an efficient iterative
algorithm for determining precoding matrices of all the users.
Our numerical results demonstrate  that the proposed iterative algorithm achieves
considerably higher sum-rate under practical QAM inputs than other known methods.
\end{abstract}



\section{Introduction}
The recent decade has witnessed the widespread application of
multiple-input multiple-output (MIMO) wireless communication systems
because of their superb spectral efficiency and link reliability \cite{Telar,Foschini,Gao}.
However, potential benefits of MIMO systems are often hampered by the omni-present interference
in typical scenarios of wireless networks \cite{Gesbert2010JSAC,Park2011,Peters2012TWC,Wu2012TWC_2}.
For this reason, considerable research interests have focused on MIMO interference channels recently.
Unlike the general tendency to assume Gaussian signal inputs in such works, our paper here
investigates linear precoding of $K$-user MIMO interference channel
systems for {\em finite alphabet input signals}.

Although there have been substantial progresses in information-theoretical analysis of Gaussian interference channels
in some special cases \cite{Sato1981IT,Costa1987IT,Sason2004IT}, the fundamental limits of the interference channels
still remain unresolved in general \cite{Etkin2008IT}.  A more recent signal multiplexing approach,
called interference alignment (IA), has shown that the degrees of freedom (DOF) of the interference channels may be achieved
in high signal-to-noise ratio (SNR) region \cite{Cadambe2008IT}. The IA concept has been extended to
MIMO interference channel systems \cite{Gou2008,Makouei2011TSP}.
Furthermore, linear precoding designs maximizing the weighted sum-rate (WSR) performance of the MIMO interference channel systems
in moderate SNR region can be found in \cite{Sung2010TWC}.

Nevertheless, the aforementioned information-theoretic works attempt to optimize the
performance of interference channel systems by relying on the convenience of
Gaussian input assumption.
Despite the information-theoretic optimality of Gaussian inputs,
practical communication systems rarely transmit Gaussian signals.
It is well known that practical signals usually are generated from
finite discrete constellation
sets such as phase-shift keying (PSK), pulse-amplitude modulation (PAM), or quadrature amplitude modulation (QAM).  MIMO precoders
designed for Gaussian input assumption may often lead to a considerable performance loss when
applied haphazardly to practical systems with finite alphabet signaling.
We take point-to-point case as an example.
For QPSK modulation, as illustrated in Figure 2 and Figure 4 in \cite{Xiao2011TSP},  the performance gaps between
the finite alphabet input design and the Gaussian input design  are more than $15$ dB at $0.75$ coding rate.

Thus, transmitter designs for optimizing the constellation-constrained mutual information
appear more plausible in practice. Such problems in the context of
point-to-point communication scenarios
\cite{Lozano2006TIT,Perez-Cruz2010TIT,Xiao2011TSP,Mohammed2011TIT} have been
recently studied.
Precoding designs in  multiple access channels,  broadcast channels, relay channels, and wiretap channels have also appeared
 \cite{Harshan2010TIT,Wang2011,Deshpande2009,Wu2012TWC,Zeng2011,Wu2012TVT}. In \cite{Abhinav2011},
 a transmit precoding design for two user single-input single-output (SISO)
 strong interference channels with finite alphabet inputs was proposed. It was suggested
 in \cite{Abhinav2011} to design the precoders by an exhaust search for the optimal
 rotation of the signal constellation at the second user side.

To the best of our knowledge, not much has been published on the MIMO interference channel
systems under finite alphabet constraints.  This paper considers the linear precoder designs to maximize the WSR
of the MIMO interference channel systems under finite alphabet constraints of channel inputs.
Our results differ significantly from previous finite alphabet research work since the existing results
on the MIMO multiuser systems with finite alphabet inputs \cite{Wang2011,Wu2012TWC} cannot be directly applied to
the MIMO interference channel systems. Besides, new insights of the precoding designs over MIMO interference channels
in asymptotic SNR regions are revealed. Moreover, a novel receiver structure exploiting the finite discrete constellation
set is provided.  The specific contributions of this paper can be
summarized as follows. Based on the mutual information between the channel input and output, we first derive an
exact expressions of the
achievable rate of each user in the MIMO interference channel systems
with finite alphabet inputs. The characterization is
applicable to $K-$users with generic antenna configurations.
We then find a near-optimal transmission strategy in low SNR region,
where each user performs beamforming in the strongest eigenmode of its channel matrix.
This transmission design coincides with the low-SNR optimal precoding policy in the point-to-point MIMO scenarios and
also for Gaussian inputs.  At high SNR, however, the precoding design with finite alphabet inputs
is significantly different from that with Gaussian inputs.
For Gaussian input, the optimal IA technique \cite{Gou2008,Sung2010TWC}
constructs the precoder of each transmitter by
properly aligning interference of each receiver to a partial signal space.
The remaining interference-free space is utilized for signal transmissions.
However, when replaced with finite alphabet inputs, the IA technique may lead to
a serious performance loss\footnote{{For Gaussian input case, it is well known that the IA technique achieves the optimal DOF in high SNR region.
The high SNR analysis in our paper  reveals that this conclusion fails to hold for finite alphabet inputs theoretically.
We admit that even for Gaussian input, the IA technique may still result in gaps from the actual rates at finite SNR levels. Therefore,
the linear precoding design over interference channels with Gaussian input for arbitrary SNR values \cite{Sung2010TWC} will also be simulated and compared
in order to report actual performance gaps at finite SNR levels.}}.  This is because finite alphabet inputs lead to saturated mutual
information.  Thus, allocating more power to signals of high SNR beyond saturation
does not further increase mutual information.
This means that using only a partial  signal space for transmitting
signals will result in a constant sum-rate loss in finite alphabet input scenarios.
For example, if each transmitter has
$N_T$ antenna and each receiver has $N_R$ antenna,
we prove that the IA technique will result in $(\eta + 1)^{-1}\sum\nolimits_{j = 1}^K {\log _2 M_j }$ b/s/Hz sum-rate loss
at high SNR, where $M_j$ denotes the size of the modulation set and $\eta$ depends on the ratio of $N_R$ and $N_T$.

As a consequence, instead of the IA, we propose a novel and concise power allocation scheme exploiting
the full signal space for signal transmission at high SNR. Our scheme applies to arbitrary number of
users and arbitrary antenna configurations. We prove that the proposed scheme
can effectively combat interference and achieve the saturated sum-rate
$\sum\nolimits_{j = 1}^K {\log _2 M_j }$ b/s/Hz at high SNR.  More generally,
due to the non-convexity of the WSR with respect to the precoding matrices,  we
derive a set of  necessary conditions
for the optimal precoding matrices through Karush-Kuhn-Tucker (KKT) analysis and further propose
an iterative algorithm using gradient descent and backtracking line search
for finding the optimal
precoders.  Numerical results show that the proposed algorithm converges within several steps and
achieves significant WSR gains over the conventional iterative designs.
In addition to WSR, the coded bit error rate (BER) of the MIMO interference channel systems is another important performance evaluation criterion in
practice. To further examine the robust performance of the proposed design, we present iterative transceiver systems for the MIMO interference channels
by deploying
low-density parity-check (LDPC) encoders and the proposed precoders at the transmitters, and the soft maximum \textit{a posteriori} (MAP) multiuser
detectors and LDPC decoders at the receivers.
Simulations show that the proposed precoding design achieves substantial coded BER improvement through
the iterative decoding and detection operations.
Furthermore, recognizing interference signals as finite alphabet inputs rather than colored Gaussian noise,
we designed a novel detector structure that provides additional coded BER gains.

The following notations are adopted throughout the paper:  Column vectors are represented by lower-case bold-face letters,
and matrices are represented using upper-case bold-face letters. Superscripts $(\cdot)^{T}$, $(\cdot)^{*}$, and $(\cdot)^{H}$
represent the matrix/vector transpose, conjugate, and conjugate-transpose operations, respectively. We let $ \left\| {\mathbf{x}} \right\|$ and $
\left\| {\mathbf{X}} \right\|_F$ denote the Euclidean norm of vector ${\mathbf{x}}$ and the Frobenius norm of matrix $\mathbf{X}$, respectively.
$\mathbb{C}$ denotes the complex field.
${\mathbf{I}}_M$ denotes an $M \times M$ identity matrix (sometimes without using subscript $M$)
and  $E_{V}$ represents the expectation of random variable (scalar, vector, or matrix) $V$.

\section{System Model and Existing Results of MIMO Interference Channel With Gaussian Inputs}\label{sec:model}

\begin{figure}[!h]
\centering
\psfrag{1}{\tiny $1$}
\psfrag{2}{\tiny $2$}
\psfrag{K}{\tiny $K$}
\psfrag{x_1}{\scriptsize
 $\mathbf{x}_1$}
    \psfrag{x_2}{\scriptsize
 $\mathbf{x}_2$}
    \psfrag{x_K}{\scriptsize
 $\mathbf{x}_{\tiny {K}}$}
   \psfrag{G_1}{\scriptsize
 $\mathbf{G}_1$}
   \psfrag{G_2}{\scriptsize
 $\mathbf{G}_2$}
 \psfrag{G_K}{\scriptsize
 $\mathbf{G}_K$}
  \psfrag{{H_ij}}{\scriptsize
 $\mathbf{H}_{ij}$}
  \psfrag{x_1_out}{\scriptsize
 $\mathbf{x}_{1,\rm{out}}$}
    \psfrag{x_2_out}{\scriptsize
 $\mathbf{x}_{2,\rm{out}}$}
    \psfrag{x_K_out}{\scriptsize
 $\mathbf{x}_{K,\rm{out}}$}
    \includegraphics[width = 0.45\textwidth]{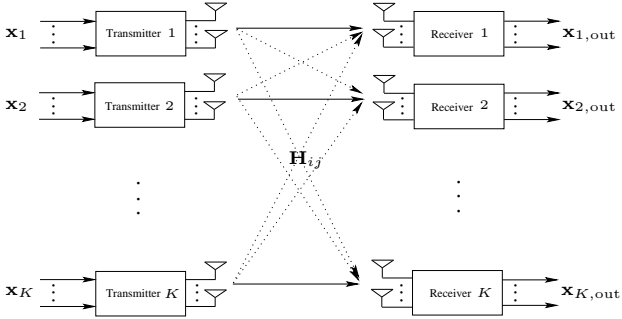}
    \caption{$K$-user interference channels with multiple antennas.}
\label{fig:model}
\end{figure}

We consider a $K$-user interference channel system in which
each transmitter has dedicated information for its intended receiver and generates
co-channel interference to other receivers as illustrated in Figure \ref{fig:model}.
Suppose the $i$-th transmitter has $N_{t_i}$ antennas and
the $j$-th receiver  has $N_{r_j}$ antennas for every $i,j = 1,2,\cdots,K$. In addition,
we assume no collaboration among transmitters or
receivers.  Then, the received signal ${\bf{y}}_j \in \mathbb{C}^{N_{r_j} \times 1}$ observed at the $j$-th receiver can be described as
\begin{equation}\label{eqn:model}
{\bf{y}}_j  = {\bf{H}}_{jj} {\bf{G}}_j {\mathbf{x}}_j  + \sum\limits_{i = 1,i \ne j}^K {{\bf{H}}_{ji} {\bf{G}}_i {\bf{x}}_i }
 + {\bf{n}}_j,\  j = 1,2, \cdots, K
\end{equation}
where ${\mathbf{x}}_j \in \mathbb{C}^{N_{t_j} \times 1}$ and ${\mathbf{n}}_j \in \mathbb{C}^{N_{r_j} \times 1}$
denote the zero-mean transmitted information vector and noise vector for the $j$-th user
with covariance matrices $\mathbf{I}$ and  $ \sigma^2  \mathbf{I}$, respectively. Also, ${\mathbf{G}}_j$
represents the linear precoding matrix for the  $j$-th user, and ${\mathbf{H}}_{ji}  \in \mathbb{C}^{N_{r_j} \times N_{t_i}}$ stands for the channel response matrix between the $i$-th transmitter
and the $j$-th receiver, which is normalized
as $\tr\left( {{\mathbf{H}}_{ji} {\mathbf{H}}_{ji}^H } \right) = N_{r_j}$. Here we make the common assumption (e.g., as in \cite{Cadambe2008IT,Sung2010TWC}, among others) that the channel state information is globally available, i.e., each transmitter has access to perfect channel knowledge of all users. Moreover, we assume the precoding matrix does not increase the transmission power. Henceforth, we have the power constraint
\begin{equation}\label{eqn:power_constraint}
\tr\left( {E\left[ {{\bf{G}}_j {\bf{x}}_j {\bf{x}}_j^H {\bf{G}}_j^H } \right]} \right) = \tr\left( {{\bf{G}}_j
{\bf{G}}_j^H } \right) \le P_j, \ j = 1,2,\cdots,K
\end{equation}
where $P_j$ denotes the maximum average transmitted power at the $j$-th transmitter.

We now briefly review the existing linear precoding design in the MIMO interference channel systems based on Gaussian input assumption.
When ${\mathbf{x}}_j \sim {\cal CN}(0, \; {\bf{I}}_{N_{t_j}}), j = 1,2,\cdots,K$, the achievable rate for the $j$-th user  is given by \cite{Sung2010TWC}
\begin{equation}\label{eqn:Rj}
\begin{array}{l}
R_j  = \log_2 \det \left( \sigma^2 {{\bf{I}} + \sum\limits_{i = 1}^K {{\bf{H}}_{ji}
{\bf{G}}_i {\bf{G}}_i^H {\bf{H}}_{ji}^H } } \right) \\
\qquad \qquad - \log_2 \det
\left( {\sigma^2 {\bf{I}} + \sum\limits_{i = 1,i \ne j}^K {{\bf{H}}_{ji} {\bf{G}}_i {\bf{G}}_i^H {\bf{H}}_{ji}^H } } \right) .
\end{array}
\end{equation}
Therefore, the optimum precoding matrices for maximizing the WSR can be expressed as
\begin{equation}\label{eqn:max_RWSR}
\left\{ {{\bf{G}}_1^* , {\bf{G}}_2^*,\cdots ,{\bf{G}}_K^* } \right\} = \mathop {\arg \max }\limits_{{\bf{G}}_1 ,{\bf{G}}_2, \cdots ,{\bf{G}}_K } R_{{\rm{wsum}}}
\end{equation}
where
\begin{equation}\label{eqn:RWSR}
\begin{array}{l}
R_{{\rm{wsum}}}  = \sum\limits_{j = 1}^K {\mu _j \log_2 \det \left(\sigma^2 {{\bf{I}} + \sum\limits_{i = 1}^K {{\bf{H}}_{ji} {\bf{G}}_i {\bf{G}}_i^H {\bf{H}}_{ji}^H } } \right)} \\
 \qquad \quad - \sum\limits_{j = 1}^K {\mu _j \log_2 \det \left(\sigma^2 {{\bf{I}} + \sum\limits_{i = 1,i \ne j}^K {{\bf{H}}_{ji} {\bf{G}}_i {\bf{G}}_i^H {\bf{H}}_{ji}^H } } \right)}
 \end{array}
 \end{equation}
and $\mu _j$ denotes the weighting factor of the $j$-th user, $j = 1,2,\cdots, K$, where  $\sum\nolimits_{j = 1}^K {\mu _j  = K}$.  Then, by exploiting the matrix derivative results in \cite{Petersen}, along with the complex matrix differentiation conclusions in \cite{Hjorungnes2011}, we can calculate the gradient of the WSR with respect to
${\bf{G}}_k$ as\footnote{We note that the derived gradient form here is a little different from \cite[Eq. (38)]{Sung2010TWC}. This is because we rewrite the WSR expression in \cite[Eq. (35)]{Sung2010TWC} only for notation simplicity. Mathematically, they are equivalent.
}
\begin{equation}\label{eqn:gradient}
\begin{array}{l}
 \nabla _{{\bf{G}}_k } R_{{\rm{wsum}}}  =  \\
 \log_2 e \left[\sum\limits_{j = 1}^K {\mu _j {\bf{H}}_{jk}^H \left(\sigma^2 {{\bf{I}} + \sum\limits_{i = 1}^K {{\bf{H}}_{ji} {\bf{G}}_i {\bf{G}}_i^H {\bf{H}}_{ji}^H } } \right)^{ - 1} } {\bf{H}}_{jk} {\bf{G}}_k  \right. \\
  - \!\! \left. \sum\limits_{j = 1,j \ne k}^K {\mu _j {\bf{H}}_{jk}^H \left(\sigma^2 {{\bf{I}} \! + \! \! \sum\limits_{i = 1,i \ne j}^K {{\bf{H}}_{ji} \! {\bf{G}}_i \! {\bf{G}}_i^H \! {\bf{H}}_{ji}^H } } \right)^{ - 1} \! {\bf{H}}_{jk}\! {\bf{G}}_k } \! \right] \! , \\
 \qquad \qquad \qquad \qquad \qquad \qquad \qquad \qquad \quad  k = 1,2,\cdots, K. \\
 \end{array}
 \end{equation}
Given the derived gradient expression, an iterative algorithm was proposed in \cite{Sung2010TWC} to solve the optimal precoders
$\left\{ {{\bf{G}}_1^* ,{\bf{G}}_2^* , \cdots , } \right.$ $ \left. {{\bf{G}}_K^* } \right\}$.

\section{Linear Precoding Design With Finite-Alphabet Inputs}\label{sec:precoder}
In practical communications, ${\mathbf{x}}_k$ is generated as equi-probably from discrete constellations
(e.g. $M_c$-ary PSK,  PAM, or QAM). In this section, we discuss the linear precoding design to maximize the WSR under the practical
finite alphabet constraints.

\subsection{Achievable Rate of Each User}
We assume ${\mathbf{x}}_j$ comes from the constellation set $S_j$ with cardinality $Q_j$,
 and define $M_j = Q_j^{N_{t_j}}$, $j = 1,2,\cdots,K$. Define $I_j$ as the $N_{t_j}$ product
 space of  $Q_j$ for the $j$-th user and  let ${\mathbf{x}}_{j,p}$ denote the $p$-th element
 in constellation set $I_j$, where $p = 1,2,\cdots,M_j$, $j = 1,2,\cdots,K$. Then, we have the following results.

\begin{prop}\label{prop:ach_rate}
Let the channel noise
${\bf{n}} \sim \mathcal{CN}({\bf{0}}, \sigma^2 {\bf{I}})$.
When the discrete input data ${\bf{x}}_j$ of the $j$-th user in the MIMO interference channel model (\ref{eqn:model}) is
independent and uniformly distributed over constellation set
$S_j$,  the achievable rate for the $j$-th user ($j = 1,2, \cdots ,K$) can be expressed as
\begin{equation} \label{eqn:achievable_Rj}
\begin{array}{l}
R_{j,\,\rm{finite}}  = \log_2 M_j - \frac{1}{{\prod\nolimits_{i = 1}^K {M_i }}}\sum\limits_{m_1  = 1}^{M_1} \sum\limits_{m_2  = 1}^{M_2} \\
\qquad \qquad { \cdots \sum\limits_{m_K  = 1}^{M_K}
 {E_{\bf{n}} \log_2 \left[ {\frac{{H_{1,j} \left( {m_1 ,m_2, \cdots ,m_K, {\mathbf{n}}} \right)}}
 {{H_{2,j} \left( {m_1 , \cdots ,m_{j - 1} ,m_{j + 1}  \cdots ,m_K ,{\mathbf{n}} } \right)}}} \right]} }
  \end{array}
\end{equation}%
where
\begin{equation} \label{eqn:achievable_H_1}
\begin{array}{l}
H_{1,j} \left( {m_1 ,m_2, \cdots ,m_K },{\bf{n}}  \right)  \\
=  \sum\limits_{n_1  = 1}^{M_1}  \sum\limits_{n_2  = 1}^{M_2}{ \! \! \cdots \! \! \sum\limits_{n_K  = 1}^{M_K} {\exp \left(\! { - \frac{{\left\| {\left( {\sum\limits_{i = 1}^K {{\bf{H}}_{ji} {\bf{G}}_i \left( {{\bf{x}}_{i,m_i }  - {\bf{x}}_{i,n_i } } \right)} } \right) + {\bf{n}}} \right\|^2 }}{{{\displaystyle\sigma} ^2 }}} \! \right)} }
 \end{array}
\end{equation}%
\begin{equation} \label{eqn:achievable_H_2}
\begin{array}{l}
 H_{2,j} \left( {m_1 , \cdots ,m_{j - 1} ,m_{j + 1} , \cdots ,m_K ,{\bf{n}} } \right) =  \sum\limits_{n_1  = 1}^{M_1 }  \cdots \sum\limits_{n_{j - 1}  = 1}^{M_{j - 1} } \\
 {\sum\limits_{n_{j + 1}  = 1}^{M_{j + 1} } { \cdots \sum\limits_{n_K  = 1}^{M_K } {\exp \left( { - \frac{{\left\| {\left( {\sum\limits_{i = 1,i \ne j}^K {{\bf{H}}_{ji} {\bf{G}}_i \left( {{\bf{x}}_{i,m_i }  - {\bf{x}}_{i,n_i } } \right)} } \right) + {\bf{n}}} \right\|^2 }}{{{\displaystyle\sigma} ^2 }}} \right)} } }  \\
 \end{array}.
 \end{equation}%
\begin{proof}
See Appendix \ref{sec:proof_ach_rate}.
\end{proof}
\end{prop}

The results in Proposition \ref{prop:ach_rate} are general for an arbitrary number of users and arbitrary
antenna configurations. For two users, the achievable rate of $R_1$ and $R_2$
can be simplified as in (\ref{R_1}) and (\ref{R_2}) at the top
of the next page.
\begin{figure*}[!ht]
\begin{equation}\label{R_1}
\begin{array}{l}
 R_{1,\, \rm{finite}}  = \log_2 M_1
 + \frac{1}{M_2}\sum\limits_{m_2  = 1}^{M_2} {E_{\bf{n}} \log_2 } \sum\limits_{n_2  = 1}^{M_2} {\exp \left( { - \frac{{\left\| {{\bf{H}}_{12} {\bf{G}}_2 \left( {{\bf{x}}_{2,m_2 }
 - {\bf{x}}_{2,n_2 } } \right) + {\bf{n}}} \right\|^2 }}{{{\displaystyle \sigma}^2 }}} \right)}  \\
 \qquad \qquad \quad  - \frac{1}{{M_1 M_2 }}\sum\limits_{m_1  = 1}^{M_1} \sum\limits_{m_2  = 1}^{M_2} E_{\bf{n}} \log_2 \sum\limits_{n_1  = 1}^{M_1} {\sum\limits_{n_2  = 1}^{M_2} {\exp \left( { - \frac{{\left\| {  {\sum\limits_{i = 1}^2 {{\bf{H}}_{1i} {\bf{G}}_i \left( {{\bf{x}}_{i,m_i }  - {\bf{x}}_{i,n_i } } \right)} }  + {\bf{n}}} \right\|^2 }}{{{\displaystyle \sigma} ^2 }}} \right)} }    \\
 \end{array}
\end{equation}
\begin{equation}\label{R_2}
\begin{array}{l}
 R_{2,\, \rm{finite}}  = \log_2 M_2 + \frac{1}{M_1}\sum\limits_{m_1  = 1}^{M_1} {E_{\bf{n}} \log_2 } \sum\limits_{n_1  = 1}^{M_1} {\exp \left( { - \frac{{\left\| {{\bf{H}}_{21} {\bf{G}}_1 \left( {{\bf{x}}_{1,m_1 }  - {\bf{x}}_{1,n_1 } } \right) + {\bf{n}}} \right\|^2 }}{{{\displaystyle \sigma} ^2 }}} \right)}  \\
\qquad \qquad \quad  - \frac{1}{{M_1 M_2 }}\sum\limits_{m_1  = 1}^{M_1} {\sum\limits_{m_2  = 1}^{M_2} {E_{\bf{n}} \log_2 \sum\limits_{n_1  = 1}^{M_1} {\sum\limits_{n_2  = 1}^{M_2} {\exp \left( { - \frac{{\left\|   {\sum\limits_{i = 1}^2 {{\bf{H}}_{2i} {{\bf{G}}_i \left( {{\bf{x}}_{i,m_i }  - {\bf{x}}_{i,n_i } } \right)} }  + {\bf{n}}} \right\|^2 }}{{{\displaystyle \sigma} ^2 }}} \right)} } }} . \\
 \end{array}
\end{equation}
 \hrulefill
\vspace*{4pt}
\end{figure*}

{In practice,  a great amount of wireless systems may work
at low power, especially for the wireless units operated by batteries.
For instance, it was reported in \cite{Bender2000CM,3GPP2001} that $40\%$ of the geographical locations
undergo receiver SNR levels below $0$ dB. In addition,
a  key objective in future digital communication systems, the energy-efficient requirement
necessitates the operation in low SNR region. It was indicated in \cite{Verdu2002TIT} that the energy
efficiency enhances as one operates in low SNR region, and
the minimum bit energy is achieved as SNR vanishes. Therefore, next we present
closed form expressions for a near-optimal transmit strategy in low SNR region.
}

\begin{prop}\label{prop:ach_rate_low}
For the interference channel model (\ref{eqn:model}), a near-optimal transmit precoding design in low SNR region ($\sigma^2 \rightarrow \infty$)
is given by
\begin{equation}\label{Precoding_low}
{\bf{G}}_j  = \sqrt{P_j} \left[ {{\bf{v}}_{\max ,j} \quad  {\mathbf{0}}} \right], \quad j = 1,2,\cdots,K
\end{equation}
where ${\bf{v}}_{\max ,j}$ is the eigenvector corresponds to the largest eigenvalue
of the matrix ${\bf{H}}_{jj}^H {\bf{H}}_{jj}$.
\begin{proof}
See Appendix \ref{sec:proof_ach_rate_low}.
\end{proof}
\end{prop}

Proposition \ref{prop:ach_rate_low} suggests that for low SNR, each user should perform
beamforming along its own channel response matrix while ignoring
the impact of interference from other users.
This is actually in agreement with intuition, since, when the SNR is low,
interferences become negligible such that
each user can design its transmission precoder accordingly to
its own channel state information.
 Thus, the precoding structure in
(\ref{Precoding_low}) directly yields the
 low-SNR optimal precoding policy in  point-to-point ($K = 1$) MIMO scenarios
\cite{Gao,Lozano2006TIT}.
In addition, theoretically, we have the following corollary.
\begin{corollary}\label{coro:low_snr}
For the interference channel model (\ref{eqn:model}) with Gaussian inputs,
the linear precoding design in Proposition \ref{prop:ach_rate_low} is optimum in low SNR region.
\begin{proof}
See Appendix \ref{sec:Gaussian_low}.
\end{proof}
\end{corollary}

Here we define ${\bf{a}}_{i,j,m_i ,n_i }  = {\bf{H}}_{ji} \left( {{\bf{x}}_{i,m_i }  - {\bf{x}}_{i,n_i } } \right)$,
$i = 1,2, \cdots ,K$; $j = 1,2, \cdots ,K$; $m_i  = 1,2, \cdots ,M_i$; $n_i  = 1,2, \cdots ,M_i$.
Let $a_{i,j,m_i ,n_i ,t}$ be the $t$th element of vector ${\bf{a}}_{i,j,m_i,n_i}$, $t = 1,2,\cdots,N_{r_j}$.
For a fixed $i$, we assume there are $T_i$  distinct $\left|{{a}}_{i,j,m_i ,n_i, t}\right| > 0$,
which we denote as $b_{i,l} ,l = 1,2, \cdots ,T_i $.  We define
\begin{align}
\omega _{i,\min } & = \min \left\{ {b_{i,l} } \right\}, \ l = 1,2, \cdots ,T_i \\
\omega _{i,\max } & = \max \left\{ {b_{i,l} } \right\}, \ l = 1,2, \cdots ,T_i .
\end{align}

Next, we give an optimal transmit precoding in high SNR region as follows:
\begin{prop}\label{prop:ach_rate_high}
For the interference channel model (\ref{eqn:model}), a transmit precoding design achieving
optimum performance
in high SNR region ($\sigma^2 \rightarrow 0$) is given by
\begin{equation}\label{Precoding_high}
{\bf{G}}_i  = \sqrt {\frac{{\varepsilon _i P_i }}{{N_{t_i} }}} {\bf{I}}_{N_{t_i} } , \quad i = 1,2, \cdots ,K
\end{equation}
where $\varepsilon _1 = 1$ and $ 0 < \varepsilon _i \leq 1, \ i = 2,3,\cdots, K$ satisfy the following conditions
\begin{equation}\label{Precoding_high_2}
\sqrt { \frac {\varepsilon _i P_i }{{N_{t_i} }}} \omega _{i,\min }  > \sum\limits_{q = i + 1}^K  \sqrt{\frac {\varepsilon _q P_q }{{N_{t_q} }}}  {\omega _{q,\max }} , \ i = 1,2,\cdots,K.
\end{equation}
Based on the precoding design in (\ref{Precoding_high}), the sum-rate of all the receivers in model (\ref{eqn:model}) at high SNR is given by
\begin{equation}\label{Precoding_high_sum}
R_{\rm sum}^\infty   = \sum\limits_{j = 1}^K {\log _2 M_j }.
\end{equation}
\begin{proof}
See Appendix {\ref{sec:proof_ach_rate_high}}.
\end{proof}
\end{prop}

Proposition \ref{prop:ach_rate_high} implies that for finite alphabet input signals, a proper power allocation scheme
can effectively eliminate the effect of interference caused by other users' transmission data\footnote{
{It should be noted that the result in Proposition \ref{prop:ach_rate_high} does not simply state an intuitive fact.
To elaborate on this, we consider a simple example of 2-user SISO interference channels. Assume $h_{11} = h_{12} = 0.5$, $g_1 = g_2 =1$, and BPSK modulation is employed.
When SNR is high (the impact of the noise vanishes), if the received signal at the receiver $1$ is $y_1 = 0$, then the receiver $1$ is unable to decode whether
the transmitted signal $x_1$ is $+1$ or $-1$. It is because the probabilities of $x_1 = 1, x_2 = -1$ and $x_1 = -1, x_2 = 1$
are equally the same.  This implies that for finite alphabet input signals, even when the constellation size at each transmitter is kept fixed
as SNR grows, each receiver may still not be able to decode all the messages from all the transmitters correctly.  As a consequence, proper
transmit design as in Proposition \ref{prop:ach_rate_high} is required to achieve the optimal performance in high SNR region.
}} with generic antenna configurations. Equation (\ref{Precoding_high_sum}) indicates that the sum-rate achieved
by the proposed scheme saturates at the bound rate of the entire signal space in high SNR region.
The IA technique \cite{Gou2008,Sung2010TWC} in contrast, which only utilizes a partial interference-free signal space for transmission,
will result in a serious performance
degradation at high SNR.  Specifically, we consider an example where $N_{t_j} = N_T$, $j=1,2,\cdots,K$ and
$N_{r_j} = N_R$, $j=1,2,\cdots,K$. We define $\eta = \lfloor\frac{\max(N_T,N_R)}{\min(N_T,N_R)}\rfloor$.
To characterize the performance loss, we have the following corollary.

\begin{corollary}\label{coro:high_snr}
Consider the case where $N_{t_j} = N_T$, $j=1,2,\cdots,K$ and
$N_{r_j} = N_R$, $j=1,2,\cdots,K$.
If $K > \eta$, the average sum-rate per channel use in model (\ref{eqn:model}) via IA technique at high SNR  is given by
\begin{equation}\label{Precoding_high_sum_gaussian}
\overline{R}_{\rm sum, \, IA}^\infty   = \frac{\eta}{\eta + 1}\sum\limits_{j = 1}^K {\log _2 M_j }.
\end{equation}
\begin{proof}
See Appendix \ref{sec:proof_high_snr}.
\end{proof}
\end{corollary}

Corollary \ref{coro:high_snr} indicates that, for MIMO interference channel with finite alphabet inputs,
a constant performance loss of the IA technique  will occur at high SNR. Particularly, for the
case where $N_T = N_R$,  the IA technique will result in a $50\%$ sum-rate loss, which will be
confirmed by our numerical results in Section IV.  This departs
markedly from the precoding design under Gaussian input assumption, where the IA technique
achieves the theoretical bound of the DOF for interference channel
in high SNR region.

\subsection{Necessary Conditions of the Optimum Precoding Matrices}
Based on Proposition \ref{prop:ach_rate}, we consider the following WSR optimization problem
\begin{equation}\label{eqn:WSR_1}
\begin{array}{l}
R_{{\rm{wsum}},\,{\rm{finite}}} ({\bf{G}}_1 ,{\bf{G}}_2, \cdots, {\bf{G}}_K) = \\
\qquad \mathop {\max }\limits_{{\bf{G}}_1 , {\bf{G}}_2,\cdots, {\bf{G}}_K } \sum\limits_{j = 1}^K {\mu _j R_{j, \,{\rm{finite}}}
\left( {{\bf{G}}_1 , {\bf{G}}_2, \cdots ,{\bf{G}}_K } \right)}
\end{array}
\end{equation}
\begin{equation}\label{eqn:WSR_2}
{\rm{subject \ to \ }} {\tr\left( {{\bf{G}}_j^H {\bf{G}}_j } \right)} \leq P_j, \ j = 1,2,\cdots,K.
\end{equation}

In general, the WSR objective function $R_{{\rm{wsum}},\,{\rm{finite}}} ({\bf{G}}_1 , {\bf{G}}_2, \cdots, {\bf{G}}_K)$
in (\ref{eqn:WSR_1})
is not concave with respect to precoding matrices $\left\{{\bf{G}}_1 , {\bf{G}}_2, \cdots, {\bf{G}}_K\right\}$. Thus, a set of necessary
conditions for this optimization problem are determined, as given in the following proposition.

\begin{prop}\label{prob:nec_cod}
The optimal precoding matrices for the WSR maximization problem in (\ref{eqn:WSR_1}) and (\ref{eqn:WSR_2}), satisfies the
following conditions
\begin{equation}\label{eqn:nec_cod_1}
\begin{array}{l}
\kappa_j {\bf{G}}_j  = \frac{{ \log_2  e}}{{{\displaystyle \sigma} ^2 }}\left( {\sum\limits_{i = 1}^K {\mu _i {\bf{H}}_{ii}^H } {\bf{T}}_{1,i,j}
- \sum\limits_{i = 1,i \ne j}^K {\mu _i {\bf{H}}_{ii}^H {\bf{T}}_{2,i,j} } } \right), \\
\qquad \qquad \qquad \qquad \qquad \qquad \qquad \qquad \qquad  j = 1,2,\cdots,K
\end{array}
\end{equation}
\begin{equation}\label{eqn:nec_cod_2}
\kappa_j \left( { {{\rm{tr}}\left( {{\bf{G}}_j^H {\bf{G}}_j } \right) - P_j} } \right) = 0, \  j = 1,2,\cdots,K
\end{equation}
\begin{equation}\label{eqn:nec_cod_3}
{ {{\rm{tr}}\left( {{\bf{G}}_j^H {\bf{G}}_j } \right) - P_j} } \leq 0, \  j = 1,2,\cdots,K
\end{equation}
\begin{equation}\label{eqn:nec_cod_4}
\kappa_j \geq 0,   \  j = 1,2,\cdots,K
\end{equation}
where the $N_{r_j} \times N_{t_j}$ matrices ${\bf{T}}_{1,i,j}$ and ${\bf{T}}_{2,i,j}$ are given as in (\ref{eqn:T_1}) and (\ref{eqn:T_2})
at the top of the next page, where ${\bf{d}}_{j,m_j ,n_j }  = {{\bf{x}}_{j,m_j }  - {\bf{x}}_{j,n_j } }$.
\begin{figure*}[!ht]
\begin{equation}\label{eqn:T_1}
\begin{array}{l}
\qquad \, {\bf{T}}_{1,i,j}  =  \\
 \frac{1}{{\prod\nolimits_{t = 1}^K {M_t } }}\sum\limits_{m_1  = 1}^{M_1 } \sum\limits_{m_2  = 1}^{M_2 } { \cdots \sum\limits_{m_K  = 1}^{M_K } {E_{\bf{n}} \left[ {\frac{{\sum\limits_{n_1  = 1}^{M_1 }  \sum\limits_{n_2  = 1}^{M_2 } { \cdots \sum\limits_{n_K  = 1}^{M_K }
 {\exp \left( { - \frac{{\left\| {\sum\limits_{t = 1}^K {{\bf{H}}_{it} {\bf{G}}_t {\bf{d}}_{t,m_t ,n_t }  + {\bf{n}}} } \right\|^2 }}{{{\displaystyle \sigma} ^2 }}} \right)\left[ {\left( {\sum\limits_{t = 1}^K { {\bf{H}}_{it} {\bf{G}}_t {\bf{d}}_{t,m_t ,n_t }  + {\bf{n}}} } \right){\bf{d}}_{j,m_j ,n_j }^H } \right]} } }}{{\sum\limits_{n_1  = 1}^{M_1} \sum\limits_{n_2  = 1}^{M_2}
 { \cdots \sum\limits_{n_K  = 1}^{M_K} {\exp \left( { - \frac{{\left\| {\sum\limits_{t = 1}^K {{\bf{H}}_{it} {\bf{G}}_t {\bf{d}}_{t,m_t ,n_t }  + {\bf{n}}} } \right\|^2 }}{{{\displaystyle \sigma} ^2 }}} \right)} } }}} \right]} }  \\
 \end{array}
\end{equation}
\begin{equation}\label{eqn:T_2}
\begin{array}{l}
 {\bf{T}}_{2,i,j}  = \frac{1}{{\prod\nolimits_{t = 1,t \ne i}^K {M_t } }}\sum\limits_{m_1  = 1}^{M_1 } { \cdots \sum\limits_{m_{i - 1}  = 1}^{M_{i - 1} } {\sum\limits_{m_{i + 1}  = 1}^{M_{i + 1} } { \cdots \sum\limits_{m_K  = 1}^{M_K } {E_{\bf{n}} } } } }  \\
 \qquad \qquad \left[ {\frac{{\sum\limits_{n_1  = 1}^{M_1 } { \cdots \sum\limits_{n_{i - 1}  = 1}^{M_{i - 1} } {\sum\limits_{n_{i + 1}  = 1}^{M_{i + 1} } { \cdots \sum\limits_{n_K  = 1}^{M_K } {\exp \left( { - \frac{{\left\| {\sum\limits_{t = 1,t \ne i}^K {{\bf{H}}_{it} {\bf{G}}_t {\bf{d}}_{t,m_t ,n_t }  + {\bf{n}}} } \right\|^2 }}{{{\displaystyle \sigma} ^2 }}} \right)} } } } \left[ {\left( {\sum\limits_{t = 1,t \ne i}^K { {\bf{H}}_{it} {\bf{G}}_t {\bf{d}}_{t,m_t ,n_t }  + {\bf{n}}} } \right){\bf{d}}_{j,m_j ,n_j }^H } \right]}}{{\sum\limits_{n_1  = 1}^{M_1 } { \cdots \sum\limits_{n_{i - 1}  = 1}^{M_{i - 1} } {\sum\limits_{n_{i + 1}  = 1}^{M_{i + 1} } { \cdots \sum\limits_{n_K  = 1}^{M_K } {\exp \left( { - \frac{{\left\| {\sum\limits_{t = 1,t \ne i}^K {{\bf{H}}_{it} {\bf{G}}_t {\bf{d}}_{t,m_t ,n_t }  + {\bf{n}}} } \right\|^2 }}{{{\displaystyle \sigma} ^2 }}} \right)} } } } }}} \right]. \\
 \end{array}
\end{equation}
 \hrulefill
\vspace*{4pt}
\end{figure*}

\begin{proof}
See Appendix \ref{sec:proof_nec_cod}.
\end{proof}
\end{prop}

\subsection{Iterative Algorithm For Weighted Sum Rate Maximization}
From (\ref{eqn:nec_cod_1}), we observe that the optimal precoding matrices of different users are
mutually dependent. For this reason, here we propose a numerical algorithm to search for the joint optimization
of $({\bf{G}}_1,{\bf{G}}_2,\cdots,{\bf{G}}_K)$.
We adopt a commonly used
suboptimal approach in dealing with multi-variables optimization problems, which iteratively optimizes one
variable at a time with others fixed \cite{Serbetli2004TSP}. During each iteration,  gradient descent method
updates
the precoding matrix for a single user. Specifically, we generate the gradient descent directions
by computing the partial mutual information derivatives of the WSR (\ref{eqn:WSR_1}) with
respect to ${\bf{G}}_j, j = 1,2, \cdots, K$, which are given by the right-hand-side of (\ref{eqn:nec_cod_1}).  The backtracking
line search algorithm is incorporated for fast convergence,  where the two related parameters $\alpha$ and $\beta$ are within $\alpha \in (0,0.5) $ and $\beta \in (0,1)$ \cite{Boyd2004}. Moreover, if the obtained solution satisfies $\tr\left\{{{\mathbf{G}}}_j{{\mathbf{G}}}_j^H\right\} > P_j$, we can project $\mathbf{{G}}_j$ to the
feasible set via a normalization step\footnote{It is noted that step 7 in Algorithm 1 updates the new precoder $\mathbf{\tilde{G}}_j^{(n)}$ along its gradient decent direction.
This is an efficient approach in searching the optimal precoder $\mathbf{G}_j$.  However, this might result in a new precoder $\mathbf{\tilde{G}}_j^{(n)}$ which
does not satisfy the power constraint.  In this case, according to the conclusion in \cite{Palomar2006TIT},  the best solution is to project  $\mathbf{\tilde{G}}_j^{(n)}$ to
the boundary of the feasible set.  At the start of the iteration, the initial weighted sum-rate is low. Thus, the weighted sum-rate will  be increased
with  the new precoder $\mathbf{G}_j^{(n+1)}$ that satisfies the individual power constraint. This has been confirmed in various scenarios in \cite{Palomar2006TIT,Wang2011,Wu2012TWC,Wu2012TVT}.
After several iteration steps, the weighted sum-rate might be high and it might not be able to find a feasible new precoder $\mathbf{G}_j^{(n+1)}$ that still increases
the weighted sum-rate. Hence, in step 6, we set a condition that when $c$ is sufficiently small, which corresponds to the sufficiently small step $t$,
 the update process stops and Algorithm 1 goes to step 11 directly.}: ${{\mathbf{G}}}_j: = \sqrt{P_j} {\mathbf{{G}}}_j /\sqrt{{{\tr\left\{{\mathbf{{G}}_j{\mathbf{G}}}_j^H\right\}}}}$ \cite{Palomar2006TIT}.

\begin{alg} \label{Gradient_IA}

Gradient descent to maximize the WSR over $ {\bf{G}}_1, {\bf{G}}_2, \cdots, {\bf{G}}_K$.
\vspace*{2.5mm} \hrule \vspace*{1mm}
  \begin{enumerate}
  \item Initialize ${\bf{G}}_{j}^ {(0)}$, with ${\tr\left( {\left({{\bf{G}}_{j}^{(0)}} \right)^H {\bf{G}}_{j}^{(0)} } \right)} = P_j$, $j = 1,2,\cdots,K$. Set $n = 0$.

  \item  Compute $R_{{\rm{wsum}},\,{\rm{finite}}} ^ {(n)}  ({\bf{G}}_{1}^{(n)} ,{\bf{G}}_{2}^{(n)}, \cdots , {\bf{G}}_{K}^{(n)})$, ${\bf{T}}_{1,i,j} ^{(n)}$
  and $ {\bf{T}}_{2,i,j} ^{(n)}$, $i = 1,2,\cdots,K, j = 1,2,\cdots,K$.

  \item  Set $j: = 1$.

  \item Compute $\nabla _{{\bf{G}}_j } R_{{\rm{wsum}},\,{\rm{finite}}}
  \left( {{\bf{G}}_1^{\left( n \right)} , {\bf{G}}_2^{\left( n \right)}, \cdots ,{\bf{G}}_K^{\left( n \right)} } \right)
   = \frac{{ \log_2  e}}{{{\displaystyle \sigma} ^2 }}\left( {\sum\nolimits_{i = 1}^K {\mu _i {\bf{H}}_i^H } {\bf{T}}_{1,i,j}^{(n)}
     - \sum\nolimits_{i = 1,i \ne j}^K {\mu _i {\bf{H}}_i^H {\bf{T}}_{2,i,j} ^ {(n)} } } \right)$.

  \item Set step size $t: = 1$.

  \item Evaluate: \\
   $c = \alpha t {  \parallel \nabla _{{\bf{G}}_j } R_{{\rm{wsum}},\,{\rm{finite}}}
  \left( {{\bf{G}}_1^{\left( n \right)} ,{\bf{G}}_2^{\left( n \right)}, \cdots ,{\bf{G}}_K^{\left( n \right)} } \right)
   \parallel_F ^ 2 }$.  If $c$ is sufficiently small,  then go to step 11.

  \item Compute: \\
   $\widetilde{\bf{G}}_j^{\left( n \right)} \! \! = \! \! {\bf{G}}_j^{\left( n \right)}
   + t\nabla _{{\bf{G}}_j } R_{{\rm{wsum}},\,{\rm{finite}}} \left( \! {{\bf{G}}_1^{\left( n \right)} ,{\bf{G}}_2^{\left( n \right)},
   \! \cdots \! ,{\bf{G}}_K^{\left( n \right)} } \! \right)$.

  \item If ${\tr\left(  \left({ \widetilde{\bf{G}}_j^{\left( n \right)} }\right)^H { \widetilde{\bf{G}}_j^{\left( n \right)} } \right)} > P_j$, update
  ${\bf{G}}_j^{\left( {n + 1} \right)}  = \frac{{\sqrt {P_j } \widetilde{\bf{G}}_j^{\left( n \right)} }}{{ { {\left\| {\widetilde{\bf{G}}_j^{\left( n \right)} } \right\|_F } } }}$;
    Otherwise, $\mathbf{G}_j^{(n+1)}=\tilde{\mathbf{G}}_j^{(n)}$.

  \item Compute: \\
   $R_{{\rm{wsum}},\,{\rm{finite}}}^{(n+1)} \! ({\bf{G}}_{1}^{(n + 1)} , \! \! \cdots\!\! , {\bf{G}}_{j-1}^{(n + 1)}\!,\!{\bf{G}}_{j}^{(n + 1)}\! , \! {\bf{G}}_{j+1}^{(n)}, \! \! \cdots \! \!, {\bf{G}}_{K}^{(n)})$, set $t : = \beta t$.

  \item If $R_{{\rm{wsum}},\,{\rm{finite}}}^{(n+1)}  ({\bf{G}}_{1}^{(n + 1)} , \!\cdots\!, {\bf{G}}_{j-1}^{(n + 1)},{\bf{G}}_{j}^{(n + 1)}, {\bf{G}}_{j+1}^{(n)}, \!\cdots\!,$ ${\bf{G}}_{K}^{(n)})
   < R_{{\rm{wsum}},\,{\rm{finite}}} ^ {(n)}  ({\bf{G}}_{1}^{(n)} ,{\bf{G}}_{2}^{(n)}, \cdots, {\bf{G}}_{K}^{(n)}) + c$, go to step 6.

  \item If $j \leq K$, $j : = j + 1$, go to step 4.

  \item Set $n : = n + 1$, go to step 2 until a stopping criterion is reached.
 \vspace*{1mm} \hrule

  \end{enumerate}
\end{alg}
\null
\par

{It is important to note that Algorithm \ref{Gradient_IA} iterates over ${\bf{G}}_j, j = 1,2, \cdots, K$
in each step increasing the WSR in (\ref{eqn:WSR_1}). Expressions in (\ref{eqn:achievable_Rj})
imply that the WSR under finite alphabet constraint is upper-bounded. As a result,
Algorithm \ref{Gradient_IA} generates increasing
sequences which are upper-bounded. Thus, it is guaranteed to converge.}
Due to the non-concavity of the objective function $R_{{\rm{wsum}},\,{\rm{finite}}} ({\bf{G}}_1 , {\bf{G}}_2,\cdots, {\bf{G}}_K)$, the proposed
algorithm may only find local maxima. To mitigate the local convergence, we randomly initialize the precoding matrices multiple
times, and choose the resulting precoders with the maximum WSR performance to be the final solution \cite{Perez-Cruz2010TIT,Wang2011,Wu2012TWC}.

The complexity of Algorithm \ref{Gradient_IA} mainly depends on the computations of
$R_{{\rm{wsum}},\,{\rm{finite}}} ^ {(n)}  \left({\bf{G}}_{1}^{(n)} ,{\bf{G}}_{2}^{(n)}, \cdots , {\bf{G}}_{K}^{(n)}\right)$,
${\bf{T}}_{1,i,j} ^{(n)}$ and $ {\bf{T}}_{2,i,j} ^{(n)}$, $i = 1,2,\cdots,K, j = 1,2,\cdots,K$.
From (\ref{eqn:achievable_Rj}), (\ref{eqn:WSR_1}), (\ref{eqn:T_1}) and (\ref{eqn:T_2}), we
note that the  computations of
$R_{{\rm{wsum}},\,{\rm{finite}}} ^ {(n)}  \left({\bf{G}}_{1}^{(n)} ,{\bf{G}}_{2}^{(n)}, \cdots , {\bf{G}}_{K}^{(n)}\right)$,
${\bf{T}}_{1,i,j} ^{(n)}$ and $ {\bf{T}}_{2,i,j} ^{(n)}$ involve
summations  of all the elements in the modulation set from all users. Hence, the complexity of Algorithm \ref{Gradient_IA} grows
exponentially with $N_{\rm{Total}}$, where $N_{\rm{Total}} = \frac{1}{\log_2 e}\sum\nolimits_{j = 1}^K  N_{t_j} \log_2 Q_j  $.

\subsection{Iterative Receiver for MIMO Interference Channels}
\begin{figure}[!h]
\centering
\psfrag{1}{\tiny $1$}
\psfrag{K}{\tiny $K$}
\psfrag{u_1}{\scriptsize $\mathbf{u}_1$}
    \psfrag{b_1}{\scriptsize $\mathbf{b}_1$}
   \psfrag{s_1}{\scriptsize $\mathbf{s}_1$}
   \psfrag{x_1}{\scriptsize $\mathbf{x}_1$}
    \psfrag{Pi}{\scriptsize $\Pi$}
   \psfrag{G_1}{\scriptsize $\mathbf{G}_1$}
 \psfrag{u_K}{\scriptsize $\mathbf{u}_{\!_K}$}
    \psfrag{b_K}{\scriptsize $\mathbf{b}_{\!_K}$}
    \psfrag{s_K}{\scriptsize $\mathbf{s}_{\!_K}$}
    \psfrag{u_K}{\scriptsize $\mathbf{u}_{\!_K}$}
       \psfrag{x_K}{\scriptsize $\mathbf{x}_K$}
    \psfrag{G_K}{\scriptsize $\mathbf{G}_{\!_K}$}
    \includegraphics[width=8.5cm, totalheight=3cm]{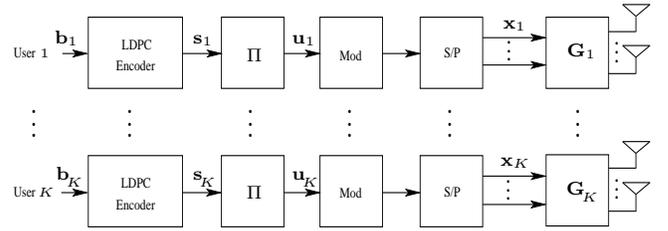}
    \caption{Transmitters of MIMO interference channels with precoding.}
\label{fig:ia_tx}
\end{figure}

\begin{figure}[!h]
\centering
\psfrag{1}{\tiny $1$}
\psfrag{K}{\tiny $K$}
    \psfrag{Pi}{\tiny
 $\Pi$}
    \psfrag{Pi_inv}{\tiny
 $\Pi^{-1}$}
    \psfrag{LE_u1}{\tiny ${\mathbf{L}}_{\!E}\!(\mathbf{u}_{_1} \!)$}
    \psfrag{LA_s1}{\tiny
 ${\mathbf{L}}_{\!A}\!(\mathbf{s}_{_1} \!)$}
    \psfrag{LD_s1}{\tiny
 ${\mathbf{L}}_{\!D}\!(\mathbf{s}_{_1} \!)$}
    \psfrag{LE_s1}{\tiny
 ${\mathbf{L}}_{\!E}\!(\mathbf{s}_{_1} \!)$}
    \psfrag{LA_u1}{\tiny
 ${\mathbf{L}}_{\!A}\!(\mathbf{u}_{_1} \!)$}
    \psfrag{b_1_head}{\tiny
 $\hat{\mathbf{b}}_{_1}$}
    \psfrag{LE_uK}{\tiny
 ${\mathbf{L}}_{\!E}\!(\mathbf{u}_{\tiny \!_K} \!)$}
    \psfrag{LA_sK}{\tiny
 ${\mathbf{L}}_{\!A}\!(\mathbf{s}_{\tiny \!_K} \!)$}
    \psfrag{LD_sK}{\tiny
 ${\mathbf{L}}_{\!D}\!(\mathbf{s}_{\tiny \!_K} \!)$}
    \psfrag{LE_sK}{\tiny
 ${\mathbf{L}}_{\!E}\!(\mathbf{s}_{\tiny \!_K} \!)$}
    \psfrag{LA_uK}{\tiny
 ${\mathbf{L}}_{\!A}\!(\mathbf{u}_{\tiny \!_K} \!)$}
    \psfrag{b_K_head}{\tiny
 $\hat{\mathbf{b}}_{\tiny \!_K}$}
\psfrag{y_1}{\tiny
 $\mathbf{y}_1$}
 \psfrag{y_K}{\tiny
 $\mathbf{y}_{\!_K}$}
    \includegraphics[width=8.5cm, totalheight=3cm]{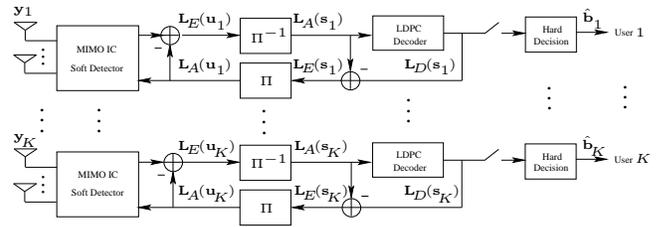}
    \caption{Iterative receivers of MIMO interference channels with precoding.}
\label{fig:ia_rx}
\end{figure}

In Algorithm \ref{Gradient_IA},  we have considered the linear precoding design  with finite alphabet inputs from the information
theoretical perspective. However, another major concern in practical communication systems is the coded BER performance.
To further examine the benefits of the proposed design,  here we present transceiver
structures in  MIMO interference channels to evaluate the coded BER performance of the obtained precoders,
which are illustrated in Figures \ref{fig:ia_tx} and \ref{fig:ia_rx}.

In Figure 2,  LDPC channel code is assumed for all the transmitters,
prior to the linear precoder designed  by Algorithm \ref{Gradient_IA}.
In  Figure 3, the turbo principle \cite{Tuchler2002Tcom} is adopted at each receiver side where the detector and the LDPC
channel decoder iteratively exchange their soft information.
$\Pi$ and $\Pi^{-1}$ in Figure 2 and Figure 3 denote the interleaver and de-interleaver respectively.
At the final iteration, hard decisions are made at each user side to obtain the estimations of the transmitted bits and the overall performance is evaluated by averaging the error rate among all the $K$ users.

Here, we choose the MAP detection method for the ``MIMO IC Soft Detector" in Figure 3, due to its near-capacity performance\footnote{Similar as in \cite{Xiao2011TSP,Wang2011,Wu2012TWC,Zeng2011}, in order to demonstrate the corresponding relationship between the mutual information performance and the coded BER performance,
``MIMO IC Soft Detector" is employed in this paper.}\cite{Hochwald2003Tcom}.
If the Gaussian input assumption is adopted, the interference for each user in (\ref{eqn:model}) can be modeled as colored Gaussian noise \cite{Park2002TVT}.  The
covariance matrices of interference-plus-noise vectors are given by
\begin{equation}\label{cov_matrix}
{\bf{C}}_j  = \sum\limits_{i = 1,i \ne j}^K {{\bf{H}}_{ji} } {\bf{G}}_i {\bf{G}}_i^H {\bf{H}}_{ji}^H  + \sigma ^2 {\bf{I}}_{N_t }, \ j = 1,2,\cdots,K.
\end{equation}
We can whiten the interference-plus-noise vectors by multiplying ${\bf{C}}_j^{-1/2}$ to obtain
\begin{equation}\label{whiten}
\widetilde{{\bf{y}} }_j = {\bf{C}}_j^{ - 1/2} {\bf{y}}_j, \ j = 1,2,\cdots,K.
\end{equation}
Then, the MAP detection method for point-to-point MIMO scenarios can be applied to evaluate the coded BER performance of each user.

However, for practical finite alphabet signals, the interference in (\ref{eqn:model}) is not Gaussian distributed. Here we provide a MAP detection
structure which does not require the Gaussian interference assumption. For the $j$-th user with received vector ${\mathbf{y}}_j$, the extrinsic LLR $L_E({{{s}}_{j,i}})$ can be expressed as \cite{Hochwald2003Tcom}
\begin{equation}\label{Ext}
\begin{array}{l}
L_E \left( {{{s}}_{j,i} } \right) = \\ \log_2 \frac{{\sum\nolimits_{{\bf{s}}_j \in {\mathbb{{S}}}_{j,i, + 1} } {p\left( {{\bf{y}}_j\left| {{\bf{x}}_j = {\rm{map}}\left( {\bf{s}}_j \right)} \right.} \right)\exp \left( {\frac{1}{2}{\bf{s}}_{j, \left[ i \right]}^T {\mathbf{L}}_A \left( {{\bf{s}}_{j,\left[ i \right]} } \right)} \right)} }}{{\sum\nolimits_{{\bf{s}}_j \in {\mathbb{{S}}}_{j,i, - 1} } {p\left( {{\bf{y}}_j\left| {{\bf{x}}_j = {\rm{map}}\left( {\bf{s}}_j \right)} \right.} \right)\exp \left( {\frac{1}{2}{\bf{s}}_{j,\left[ i \right]}^T {\mathbf{L}}_A \left( {{\bf{s}}_{j,\left[ i\right]} } \right)} \right)} }}
\end{array}
\end{equation}
where ${{s}}_{j,i}$ denotes the $i$-th bit of the $j$-th user, with $1 \leq i \leq N_{t_j} M_j$, and $1 \leq j \leq K$. The $N_{t_j} M_j \times 1$ vector ${\mathbf{s}}_j$ represents the coded bits for the $j$-th user.  ${\mathbb{S}}_{j, i, + 1}$ and ${\mathbb{S}}_{j, i, - 1}$  are the sets of $2^{N_{t_j} M_j-1}$ bit vectors  with the $i$-th element being $+ 1$ and $- 1$, respectively.  ${{\bf{s}}_{j,\left[ i \right]} }$ demonstrates the subvector of ${\bf{s}}_j$
by omitting the $i$-th elements. ${\mathbf{L}}_A \left( {{\bf{s}}_{j,\left[ i \right]} } \right)$ is the $(N_{t_j} M_j -1) \times 1$ vector with the \textit{a priori} information of
${{\bf{s}}_{j,\left[ i \right]}}$.  ${\bf{x}}_j  = {\rm{map}}\left( {{\bf{s}}_j } \right)$ denotes the modulation from the bit vector ${\bf{s}}_j$ to symbol vector ${\bf{x}}_j$. For the MIMO interference channel systems (\ref{eqn:model}), we have the likelihood function
\begin{equation}\label{li_funtion}
\begin{array}{l}
 p\left( {{\bf{y}}_j \left| {{\bf{x}}_j  = {\rm{map}}\left( {{\bf{s}}_j } \right)} \right.} \right)
 \!  =  \! \frac{1}{{\left( {\pi \sigma ^2 } \right)^{N_r } \prod\nolimits_{t = 1, t \neq j}^K {M_t } }}\sum\limits_{m_1  = 1}^{M_1 } \! \cdots  \! \sum\limits_{m_{j - 1}  = 1}^{M_{j - 1} }   \\   \sum\limits_{m_{j + 1}  = 1}^{M_{j + 1} } { \!\! \cdots  \! \! \sum\limits_{m_K  = 1}^{M_K } {\exp \left( { - \frac{{\left\| {{\bf{y}}_j  - {\bf{H}}_{jj} {\bf{G}}_j {\bf{x}}_j  - \sum\limits_{t = 1,t \ne j}^K {{\bf{H}}_{jt} {\bf{G}}_t {\bf{x}}_{t,m_t} } } \right\|^2 }}{{{\displaystyle \sigma} ^2 }}} \right)} }  .  \\
 \end{array}
\end{equation}

\section{Numerical Results}
In this section,  we provide numerical results to examine the performance of the proposed iterative optimization algorithm.
We assume equal individual power limit $P_1 = \cdots P_{K} = P$, normalized channel matrices, and the same modulation for all $K$ users.
Then, the average SNR of the MIMO interference channel systems is given by \cite{Sung2010TWC} ${\rm{SNR}} = P  /\sigma ^2 $. It is noted
that all the  channel matrices in this section are generated randomly.

First, we consider a $2$-user MIMO interference channel system with two transmit and two receive antennas for each user.
Similar to \cite{Wang2011}, for illustrative purpose, here we assume the channel matrices for two users are fixed (non-fading), which are given by
\begin{eqnarray}
{\bf{H}}_{11} =&&\hspace{-0.6cm} \left( \begin{array}{lr}
  1.2813 & 0.1578 + 0.3445j  \\
  0.1578 - 0.3445j & 0.2666  \\
\end{array} \right) \nonumber \\
 {\bf{H}}_{12} = &&\hspace{-0.6cm} \left( \begin{array}{lr}
   0.4596 & 0.0332 - 0.5936j  \\
   0.0332 + 0.5936j & 1.0401  \\
\end{array} \right) \nonumber \\
{\bf{H}}_{21}  = &&\hspace{-0.6cm} \left( \begin{array}{lr}
   0.3523   & -0.3938 + 0.3207j  \\
  -0.3938 - 0.3207j  &   1.1662   \\
\end{array} \right) \nonumber \\
{\bf{H}}_{22}  = &&\hspace{-0.6cm} \left( \begin{array}{lr}
 0.2483  & 0.3246 + 0.2157j  \\
   0.3246 - 0.2157j & 1.2785   \\
\end{array} \right). \nonumber
\end{eqnarray}

\begin{figure}[!h]
\centering
\includegraphics[width=0.5\textwidth]{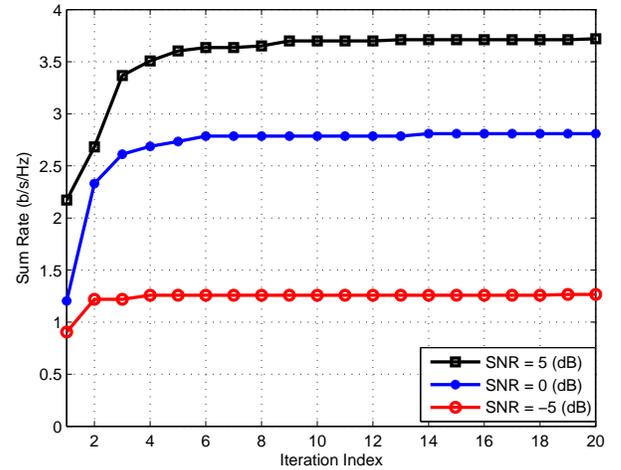}
\caption {\space\space Convergence of sum rate maximization with BPSK modulation. }
\label{converage_bpsk}
\end{figure}

\begin{figure}[!h]
\centering
\includegraphics[width=0.5\textwidth]{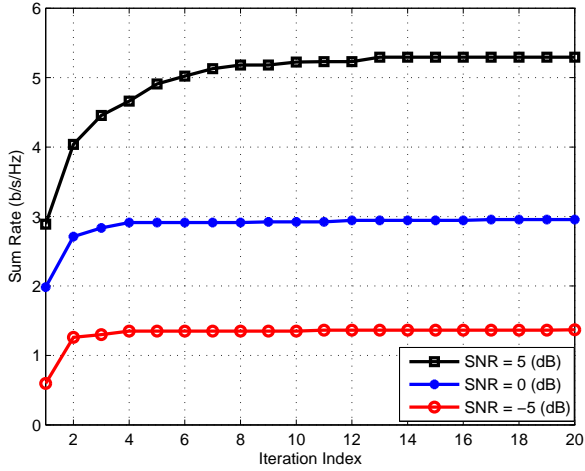}
\caption {\space\space Convergence of sum rate maximization with QPSK modulation.  }
\label{converage_qpsk}
\end{figure}

Figures \ref{converage_bpsk} and \ref{converage_qpsk} show the convergence behaviors of the proposed algorithm at different SNRs and under different modulations for the sum-rate ($\mu_1 = \mu_2 = 1$) maximization. For backtracking line search in  Algorithm \ref{Gradient_IA}, the typical parameters
are chosen as $\alpha = 0.1$ and  $\beta = 0.5$ \cite{Boyd2004}.
Statistical averages in  (\ref{eqn:achievable_Rj})  and (\ref{eqn:nec_cod_1})
are evaluated via Monte Carlo drawings of $1000$ random samples.
These figures illustrate the evolutions of the sum-rate for each iterative step.
We can see that in all the cases, the proposed algorithm converges within about 10 iteration steps.

\begin{figure}[!h]
\centering
\includegraphics[width=0.5\textwidth]{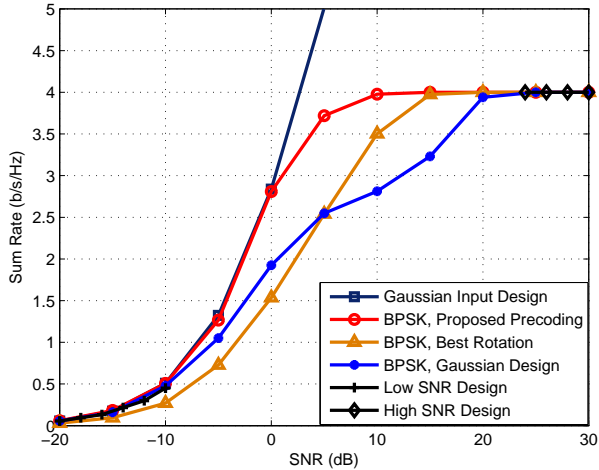}
\caption {\space\space Sum rate of $2$-user MIMO interference channels with BPSK modulation. }
\label{sum_rate_bpsk}
\end{figure}

Figure \ref{sum_rate_bpsk} compares the sum-rate performances\footnote{It is noted that practical techniques
can be employed in the random initial point selection process.
For instance, the initial precoder at current SNR level can be
generated by a weighted combination of the
final precoder obtained at previous SNR level and a randomly generated
precoder. Since the SNR step in our simulation is not large, this can
take advantage of the previously computed precoder and accelerate the
convergence speed. In general, for finite alphabet input signals,
the WSR value lies in a limited bounded space. Therefore,  2-3 times random initializations are
often sufficient to achieve a near-optimal WSR performance. } of BPSK modulation under different transmit precoding schemes.
 We extend the idea for two-user SISO
interference channels in \cite{Abhinav2011} by a numerical
search for the best rotation angle $\theta$ for the second user constellation set $e^{j\theta } {\bf{x}}_{2,j} ,j = 1, 2,\cdots ,M_2,
\theta \in [0, 2 \pi)$, which is denoted as ``BPSK, Best Rotation".  With respect to  ``BPSK, Gaussian Design", we implement the Gaussian input assumption linear precoding design by the iterative algorithm in \cite{Sung2010TWC}, and compute the finite alphabet rate of this precoding design in (\ref{eqn:achievable_Rj}).
The ``Low SNR Design",  the ``High SNR Design", and  the sum-rate achieved by Gaussian input assumption in \cite{Sung2010TWC}
are also evaluated and plotted.

From Figure \ref{sum_rate_bpsk}, we have the following observations:

\begin{itemize}
\item[1)]  The proposed precoding has BPSK system performance close to the sum-rate upper bound
achieved by Gaussian inputs when SNR is below $0$ dB.

\item[2)]    Our proposed precoding provides obvious sum rate gains
over comparison schemes throughout the tested SNR region. At the sum-rate of $3$ b/s/Hz,
the proposed precoding is about $6.5$ dB and $12$ dB
better than the ``BPSK, Best Rotation" and ``BPSK, Gaussian Design", respectively.

\item[3)] The performances of the ``Low SNR Design" and the sum-rate achieved by  Gaussian inputs
are virtually identical at low SNR.

\item[4)]   The sum-rate of the ``High SNR Design"  saturates at
$\sum \nolimits_{j = 1}^K \log_2{M_j }= 4$ b/s/Hz in high SNR region.

\end{itemize}

For BPSK modulation, the final precoding matrices obtained via Algorithm \ref{Gradient_IA} when ${\rm{SNR}} = 5$ dB
are given by
\begin{eqnarray}
 {\mathbf{G}}_1 = &&\hspace{-0.6cm}
 \left[ \begin{array}{lr}
0.5390 - 0.7978j   &  1.0204 + 0.1993j \\
  -1.0166 - 0.1732j &  0.2907 + 0.0809j \\
 \end{array} \right] \nonumber \\
{\mathbf{G}}_2 = &&\hspace{-0.6cm}
 \left[\begin{array}{lr}
  -0.0063 + 0.0404j  &  0.2802 - 0.3445j \\
  1.2232 + 0.0059j &  0.6386 - 1.0292j \\
 \end{array} \right] . \nonumber
 \end{eqnarray}

\begin{figure}[!h]
\centering
\includegraphics[width=0.5\textwidth]{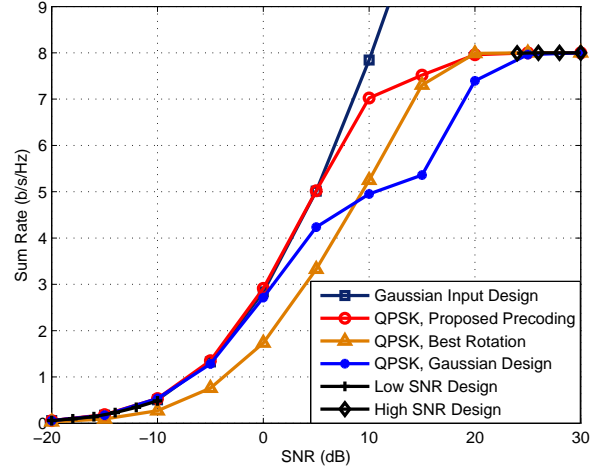}
\caption {\space\space  Sum rate of $2$-user MIMO interference channels with QPSK modulation.  }
\label{sum_rate_qpsk}
\end{figure}

The sum-rate performances under different transmit precoding schemes
with QPSK inputs are illustrated in Figure \ref{sum_rate_qpsk}.
We have similar observations
with the BPSK case, where the proposed precoding achieves higher sum-rate than other precoding schemes, and
the sum-rate performance of the ``Low SNR Design"  matches closely with the performance of the Gaussian input case in low SNR region.
At a targeted sum-rate of $6$ b/s/Hz, the performance gains of the proposed precoding over the ``QPSK, Best Rotation" and ``QPSK, Gaussian Design"
 are about $4.5$ dB and\footnote{It should be noted that as the constellation size increases, the performance gap between the finite alphabet input design
and the Gaussian input design will decrease.} $10$ dB, respectively.  Note that the sum-rate performances of the ``High SNR Design" achieve
 $ \sum \nolimits_{j = 1}^K \log_2{M_j } = 8$ b/s/Hz in high SNR region.
Furthermore, we observe that in both BPSK and QPSK cases, linear precoding design based
on Gaussian input assumption  performs almost identically as the proposed precoding in low SNR region, which
corroborates the conclusion in Corollary \ref{coro:low_snr}.

For QPSK modulation, the final precoding matrices obtained via Algorithm \ref{Gradient_IA}  when ${\rm{SNR}} = 5$ dB
are given by
\begin{eqnarray}
  {\mathbf{G}}_1 = &&\hspace{-0.6cm}  \left[ \begin{array}{lr}
   1.0523 - 0.6658j  &  0.2312 - 1.1369j \\
   0.3759 - 0.0377j  & 0.2090 - 0.2814j \\
 \end{array} \right] \nonumber \\
  {\mathbf{G}}_2 = &&\hspace{-0.6cm} \left[ \begin{array}{lr}
  -0.4825 + 0.0572j  &  0.2828 + 0.1678j \\
  -0.8290 + 1.0292j  & 1.0257 + 0.1388j \\
 \end{array} \right]. \nonumber
 \end{eqnarray}

\begin{figure}[!h]
\centering
\includegraphics[width=0.5\textwidth]{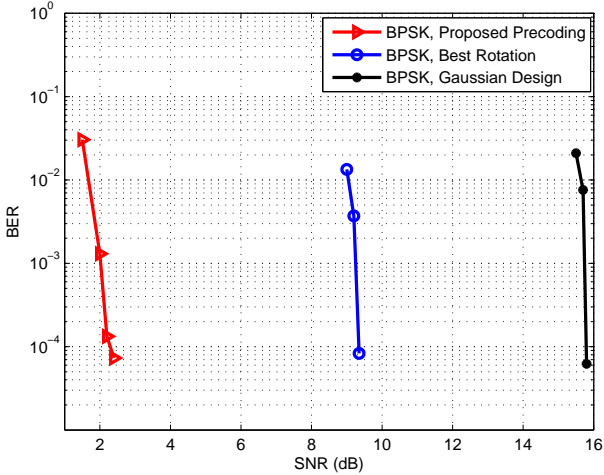}
\caption {\space\space BER of $2$-user MIMO interference channels  with BPSK modulation.  }
\label{ber_bpsk}
\end{figure}

As a more comprehensive comparison, we obtain the coded BER performances of a 2-user MIMO interference channel system illustrated in Figure \ref{fig:ia_tx} and \ref{fig:ia_rx}.
The LDPC encoder and decoder simulation package \cite{Valenti} is used, for coding rate $3/4$ and coding length $L = 9600$.
The MAP detector given in (\ref{Ext}) is employed. The number of iterations between the MAP detector and the LDPC decoder is
set to 5. Figure 8 plots the coded BER curves with different precoding schemes under BPSK modulation. We observe from Figure \ref{ber_bpsk} that,
 at a targeted BER of $10^{-4}$, the corresponding SNR gains of the proposed precoding over ``BPSK, Best Rotation"
  and ``BPSK, Gaussian Design"
  schemes are about $7$ dB and $13.5$ dB, respectively. This suggests
 that the proposed precoding is very promising
as it shows direct improvement on the coded BER performance in practical systems. Furthermore, comparing  the performances of
the proposed precoding and the Gaussian schemes in Figure \ref{sum_rate_bpsk} and Figure \ref{ber_bpsk},  we note that the SNR gains
for the coded BER are larger than the corresponding SNR gains for the sum-rate shown in Figure \ref{sum_rate_bpsk}.
The reason for this observation is that we average the
coded BER of $2$ users to generate Figure \ref{ber_bpsk}.
Therefore, the average coded BER is dominated by the user with larger coded BER.

\begin{figure}[!h]
\centering
\includegraphics[width=0.5\textwidth]{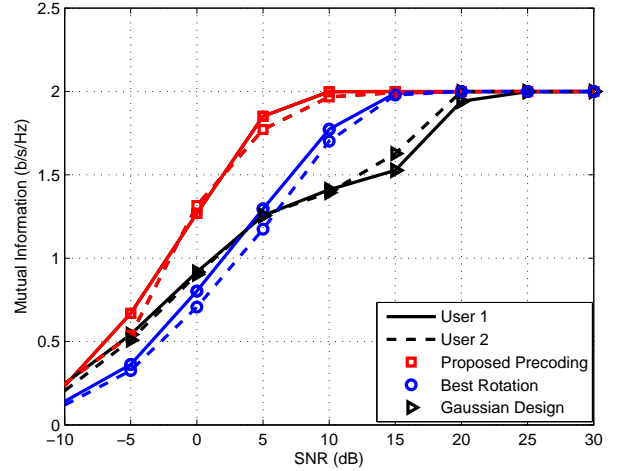}
\caption {\space\space Mutual information for individual user  with BPSK modulation. }
\label{ind_rate_bpsk}
\end{figure}

To confirm this, we present the mutual information of individual users
under different precoding schemes and BPSK modulation in Figure \ref{ind_rate_bpsk}.
It is revealed in Figure \ref{ind_rate_bpsk} that, e.g., although the sum-rate is above $3$ b/s/Hz for
the  Gaussian precoder design when ${\rm{SNR}} = 14$ dB, the achievable rate for user 1 is below $1.5$ b/s/Hz.
Therefore, the overall coded BER is
 still high for the Gaussian design in Figure \ref{ber_bpsk} when ${\rm{SNR}} = 14$ dB.
Also, we can observe from Figure \ref{ind_rate_bpsk} that the SNR gain for the user 1 of
the proposed precoding over user 1 of the Gaussian design at rate $1.5$ b/s/Hz
is about $13.5$ dB. The results correspond to the SNR gains for the coded BER in Figure \ref{ber_bpsk}.

\begin{figure}[!h]
\centering
\includegraphics[width=0.5\textwidth]{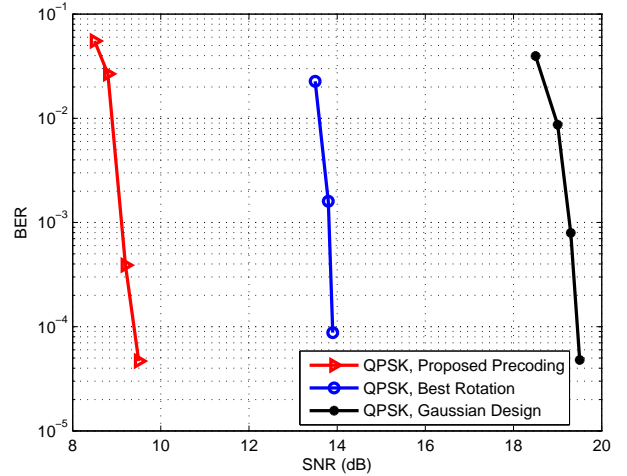}
\caption {\space\space BER of $2$-user MIMO interference channels  with QPSK modulation. }
\label{ind_ber_qpsk}
\end{figure}

\begin{figure}[!h]
\centering
\includegraphics[width=0.5\textwidth]{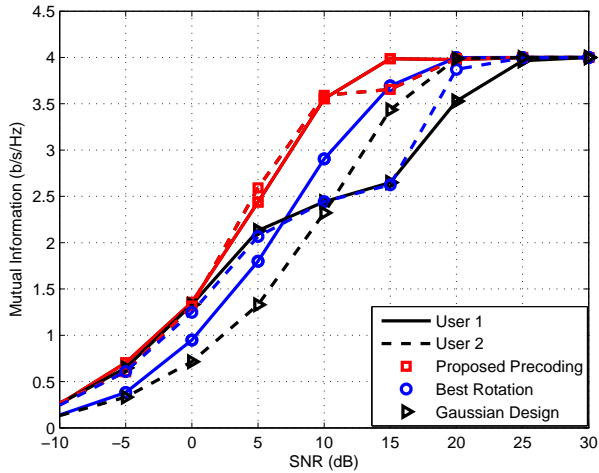}
\caption {\space\space Mutual information for individual user with QPSK modulation. }
\label{ind_rate_qpsk}
\end{figure}

Figure \ref{ind_ber_qpsk} depicts the coded BER curves with QPSK inputs.
The observations in Figure \ref{ind_ber_qpsk} are similar with the BPSK case. At a targeted coded BER
of $10^{-4}$, the proposed precoding design has $5.5$ dB and $10$ dB SNR gains over
``QPSK, Best Rotation" and ``QPSK, Gaussian Design", respectively.
Also, the coded BER results coincide  with the  mutual information results for individual users
under QPSK modulation and different precoder designs as shown in Figure \ref{ind_rate_qpsk}.

Next, the coded BER performances of using the detection method in (\ref{whiten}) (denoted as ``Gaussian Assumption Detection") and the detection method in (\ref{Ext}) (denoted as ``Proposed Detection") under different modulations  are  compared.
Simulation results are listed in Table I. The precoders obtained via Algorithm \ref{Gradient_IA} are
used at the transmitter. From the results shown in Table I, it is seen that the ``Proposed Detection" achieves lower BER than
the ``Gaussian Assumption Detection", which indicates that exploiting the interference constellation structure
can lead to additional performance gains in practical systems.

\begin{table}[!h]
\label{Table_BER}
\centering
 \captionstyle{center}
\vspace*{1mm}  \caption{BER Results Using Different Detection Methods.} \vspace*{2mm}  \label{tab:relation}
\begin{tabular}{|c|c|c|c|}
\hline
\multirow{3}{*} {\textbf{Modulation}}  &   \multirow{3}{*}{{\rm{\textbf{SNR}} \textbf{(dB)}}} &   \multirow{2}{*}{\textbf{Proposed}}  &  \textbf{Gaussian} \\
  &  &   \multirow{2}{*}{\textbf{Detection}}  &  \textbf{Assumption}    \\
 &  &  & \textbf{Detection}    \\
 \hline
 \multirow{4}{*} {\textbf{BPSK}}   &   2  &  0.0013    &   0.0114  \\
    & 2.5    &  0  &  $7.513 \times 10^{-4}$ \\
  &   3 &  0  & 0  \\
  &   3.5   &   0  &   0  \\
  \hline
  \multirow{4}{*} {\textbf{QPSK}}  &   9   & 0.0086   &  0.0780  \\
  &   10  &  0  & 0.0712  \\
  &   11 &  0  &  0.0578  \\
  &   12   &   0  &  0.0476  \\
   \hline
\end{tabular}
\end{table}

\begin{figure}[!h]
\centering
\includegraphics[width=0.5\textwidth]{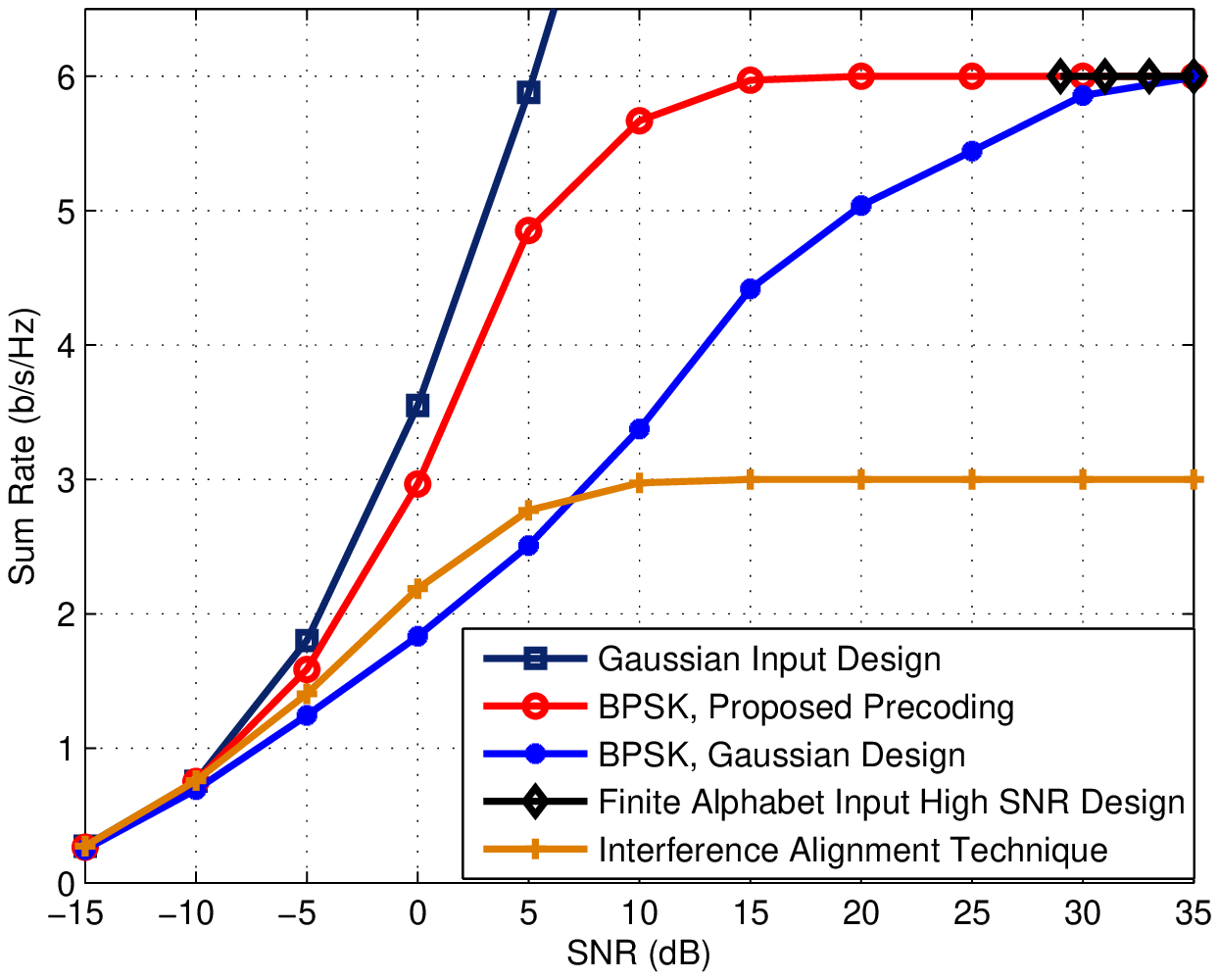}
\caption {\space\space Sum rate of $3$-user MIMO interference channels with BPSK modulation. }
\label{sum_rate_bpsk_three}
\end{figure}

\begin{figure}[!h]
\centering
\includegraphics[width=0.5\textwidth]{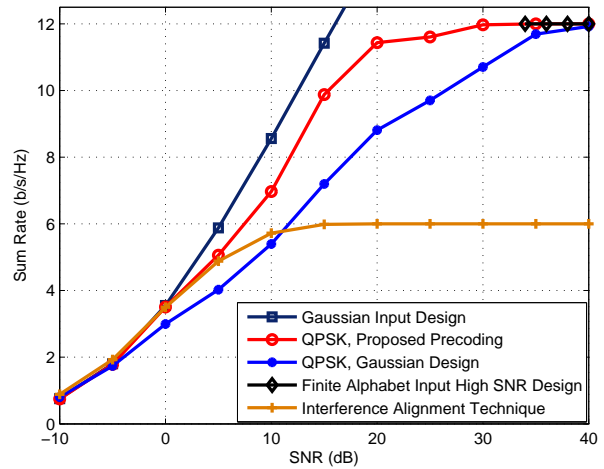}
\caption {\space\space Sum rate of $3$-user MIMO interference channels with QPSK modulation. }
\label{sum_rate_qpsk_three}
\end{figure}

Finally, we further investigate the precoding design in a $3$-user MIMO interference channel system.
The channel matrices are given by
\begin{eqnarray}
{\bf{H}}_{11}  = &&\hspace{-0.6cm} \left( \begin{array}{lr}
 -0.2279 - 0.6039j & 0.0660 + 0.6264j  \\
 -0.8774 + 0.6273j & 0.1515 - 0.0198j  \\
\end{array} \right) \nonumber \\
{\bf{H}}_{12}  = &&\hspace{-0.6cm} \left( \begin{array}{lr}
   0.1915 - 0.3442j &  - 0.1092 - 0.0798j  \\
   0.1022 + 1.2773j &  0.4246 + 0.0667j  \\
\end{array} \right) \nonumber   \\
{\bf{H}}_{13}  = &&\hspace{-0.6cm} \left( \begin{array}{lr}
   0.5759 + 0.1583j &  - 0.1092 - 0.0798j  \\
   0.1022 + 1.2773j &  0.4246 + 0.0667j  \\
\end{array} \right) \nonumber
\end{eqnarray}

\begin{eqnarray}
{\bf{H}}_{21}  = &&\hspace{-0.6cm} \left( \begin{array}{lr}
    - 0.3382 - 0.7046j &  0.6131 - 0.1955j  \\
    0.4195 + 0.2793j &  - 0.7792 + 0.3374j  \\
\end{array} \right) \nonumber \\
{\bf{H}}_{22}  = &&\hspace{-0.6cm} \left( \begin{array}{lr}
   0.4643 + 0.6778j & 0.7344 - 0.0113j  \\
   0.4052 - 0.6845j & 0.3806 - 0.0892j  \\
\end{array} \right)\nonumber \\
{\bf{H}}_{23}  = &&\hspace{-0.6cm} \left( \begin{array}{lr}
   - 0.8238 - 0.4134j & 0.5425 + 0.1126j  \\
  0.1321 + 0.2715j & 0.7267 - 0.4734j  \\
\end{array} \right) \nonumber
 \end{eqnarray}

\begin{eqnarray}
{\bf{H}}_{31}  = &&\hspace{-0.6cm} \left( \begin{array}{lr}
 - 0.9707 + 0.2271j &  - 0.4520 - 0.2644j  \\
 - 0.0265 - 0.7569j &  0.2748 - 0.2878j  \\
\end{array} \right) \nonumber \\
{\bf{H}}_{32}  =&&\hspace{-0.6cm} \left( \begin{array}{lr}
    - 0.2200 - 0.0000j &  - 0.0113 + 0.6334j  \\
    - 0.5837 - 0.1839j &  - 0.0279 - 1.0840j  \\
\end{array} \right)\nonumber \\
{\bf{H}}_{33}  =&&\hspace{-0.6cm} \left( \begin{array}{lr}
    - 0.2200 - 0.0000j &  - 0.0113 + 0.6334j  \\
    - 0.5837 - 0.1839j &  - 0.0279 - 1.0840j  \\
\end{array} \right). \nonumber
 \end{eqnarray}

Figure \ref{sum_rate_bpsk_three} compares the sum-rate performances of different precoding designs with BPSK inputs. The
sum-rates achieved by the proposed ``Finite Alphabet Input High SNR Design" and the
``Interference Alignment Technique" are also plotted.
It is observed from Figure \ref{sum_rate_bpsk_three} that the proposed precoding outperforms
other precoding  designs.  At  a
targeted sum-rate of $4.5$ b/s/Hz, our proposed precoding has about $10$ dB SNR gains over the Gaussian design.
Moreover, the sum-rate performances with QPSK inputs
are plotted in Figure \ref{sum_rate_qpsk_three}. At a
targeted sum-rate of $9$ b/s/Hz,
the performance gains of the  proposed precoding over the Gaussian design are
about $8$ dB. We note that, in both modulations, the sum-rates achieved by
the ``Finite Alphabet Input High SNR Design" saturate at $R_{\rm{sum}}^\infty   = \sum\nolimits_{j = 1}^K {\log _2 M_j }$ b/s/Hz,
and the ``Interference Alignment" has $50\%$ sum-rate losses\footnote{It is noted that each user transmits one data
stream for the IA method. Although the IA approach can align other users' interference into a signal space and eliminate them,
it is still suboptimal. This is because the mutual information with finite alphabet inputs is bounded.  Thus, allocating more power to
the already saturated signals can not further improve the mutual information.
This implies that transmitting only one data stream for each user
will result in a constant sum-rate loss in finite alphabet input scenarios. Instead, for the proposed method in Algorithm 1 and the high SNR design in Proposition 3, each user transmits two data streams. It is shown in Figure 12 and Figure 13 that  both Algorithm 1
and the proposed high SNR design in Proposition 3 can effectively combat interference and achieve the saturated sum-rate
$\sum\nolimits_{j = 1}^K {\log _2 M_j }$ b/s/Hz at high SNR.} as implied
in Corollary 2.

\section{Conclusion}
This work investigated the linear precoding design for MIMO interference
channel systems. To break away from the traditional studies based on
Gaussian input assumption, we formulated the problem of maximizing mutual
information for finite alphabet inputs.
We derived the achievable rate expression of each user.
Our analysis at the high SNR revealed that the
IA technique designed for Gaussian input case will lead to
a significant sum-rate loss due to the utilization of
partial interference-free signal space for transmission.
In light of this, we developed an efficient power allocation scheme designed for finite alphabet input scenario at high SNR.
The proposed scheme achieves the analytical upper-bound sum-rate of the entire signal space $\sum\nolimits_{j = 1}^K {\log _2 M_j }  $ b/s/Hz
under finite alphabet constraints.
More generally, we derived a set of necessary conditions for the WSR maximization precoding design
based on  Karush-Kuhn-Tucker (KKT) analysis, from which
we developed an iterative algorithm for precoding optimization. We applied
gradient descent optimization and used backtracking line
search algorithm to regulate the convergence speed.  Our tests, based on
LDPC coded QAM signals in
the MIMO interference channels established the coded BER performance gain of the obtained linear precoders.

\section*{Acknowledgment}
The authors would like to thank the editor, Prof. Tony Q. S. Quek,
and  anonymous reviewers for helpful comments and suggestions that greatly improved
the quality of the paper.

\appendices
\section{Proof of Proposition \ref{prop:ach_rate}}\label{sec:proof_ach_rate}
When the discrete input data vector ${\mathbf{x}}_k$ is
independent and uniformly distributed from the constellation set $S_j$, the \textit{a priori} probabilities of ${\mathbf{x}}_{j,p}$ is
$p({\mathbf{x}}_j = {\mathbf{x}}_{j,p}) = \frac{1}{{M_j }}$, $p = 1,2,\cdots,M_j$, $j = 1,2,\cdots,K$. Then, according to the definition
of conditional entropy \cite{Cover} and the Bayes' theorem, we can evaluate $H\left( {{\bf{y}}_j } \right)$ and  $H\left( {{\bf{y}}_j \left| {{\bf{x}}_j } \right.} \right)$
given in (\ref{eqn:H_y_3}) and (\ref{eqn:H_yx_4}) at the top of the next page.
\begin{figure*}
\begin{equation}\label{eqn:H_y_3}
\begin{array}{l}
H\left( {{\bf{y}}_j } \right) = \sum\limits_{i = 1}^K {\log_2 M_i  + \frac{1}{{\prod\nolimits_{i = 1}^K {M_i } }}}
 \sum\limits_{m_1  = 1}^{M_1 }  \sum\limits_{m_2  = 1}^{M_2 }\cdots
  {  \sum\limits_{m_K  = 1}^{M_K } {{\scalebox{1}[1.2]{$\int$}} {p\left( {{\bf{y}}_j \left| {{\bf{x}}_{1,m_1 } ,{\bf{x}}_{2,m_2 },\cdots,{\bf{x}}_{K,m_K } } \right.} \right)} } }  \\
 \qquad \qquad \qquad \qquad \times \log_2 \frac{1}{{\sum\limits_{n_1  = 1}^{M_1 } \sum\limits_{n_2  = 1}^{M_2 }
 { \cdots \sum\limits_{n_K  = 1}^{M_K } {p\left( {{\bf{y}}_j \left| {{\bf{x}}_{1,n_1 } ,{\bf{x}}_{2,n_2 },\cdots,{\bf{x}}_{K,n_K } } \right.} \right)} } }}d{\bf{y}}_j  \\
 \end{array}
\end{equation}
 \begin{equation}\label{eqn:H_yx_4}
\begin{array}{l}
H\left( {{\bf{y}}_j \left| {{\bf{x}}_j } \right.} \right)  = \sum\limits_{i = 1,i \ne j}^K {M_i }  + \frac{1}{{\prod\nolimits_{i = 1}^K {M_i } }}\sum\limits_{m_1  = 1}^{M_1 } \sum\limits_{m_2  = 1}^{M_2 } \cdots
  {  \sum\limits_{m_K  = 1}^{M_K } {{\scalebox{1}[1.5]{$\int$}} {p\left( {{\bf{y}}_j \left| {{\bf{x}}_{1,m_1 } ,{\bf{x}}_{2,m_2 },\cdots,{\bf{x}}_{K,m_K } } \right.} \right)} } } \qquad \qquad
  \qquad \qquad \, \\
\qquad  \qquad \qquad  \qquad \times \log_2 \frac{1}{{\sum\limits_{n_1  = 1}^{M_1 } { \cdots \sum\limits_{n_{j - 1}  = 1}^{M_{j - 1} } {\sum\limits_{n_{j + 1}  = 1}^{M_{j + 1} } { \cdots \sum\limits_{n_K  = 1}^{M_K }
{p\left( {{\bf{y}}_j \left| {{\bf{x}}_{j,m_j} ,{\bf{x}}_{1,n_1 },\cdots,{\bf{x}}_{j - 1,n_{k - 1} } ,{\bf{x}}_{j + 1,n_{j + 1} } ,...,{\bf{x}}_{K,n_K } } \right.} \right)} } } } }}d{\bf{y}}_j .  \\
 \end{array}
 \end{equation}
 \hrulefill
\vspace*{4pt}
\end{figure*}

From model (\ref{eqn:model}), we have
\begin{equation}\label{eqn:p1}
\begin{array}{l}
p\left( {{\bf{y}}_j \left| {{\bf{x}}_{1,m_1 } ,{\bf{x}}_{2,m_2 }, \cdots ,{\bf{x}}_{K,m_K } } \right.} \right) \\
 \qquad \qquad \qquad = \frac{1}{{\left( {\pi {\displaystyle \sigma} ^2 } \right)^{N_{r_j } } }}\exp \left( { - \frac{{\left\| {{\bf{y}}_j  - \sum\limits_{i = 1}^K {{\bf{H}}_{ji} {\bf{G}}_i {\bf{x}}_{i.m_i } } } \right\|^2 }}{{{\displaystyle \sigma} ^2 }}} \right)
\end{array}
\end{equation}

\begin{equation}\label{eqn:p2}
\begin{array}{l}
p\left( {{\bf{y}}_j \left| {{\bf{x}}_{1,n_1 } , {\bf{x}}_{2,n_2 }, \cdots ,{\bf{x}}_{K,n_K } } \right.} \right) \\
 \qquad \qquad \qquad = \frac{1}{{\left( {\pi {\displaystyle \sigma} ^2 } \right)^{N_{r_j } } }}\exp \left( { - \frac{{\left\| {{\bf{y}}_j  - \sum\limits_{i = 1}^K {{\bf{H}}_{ji} {\bf{G}}_i {\bf{x}}_{i.n_i } } } \right\|^2 }}{{{\displaystyle \sigma} ^2 }}} \right)
\end{array}
\end{equation}

\begin{equation}\label{eqn:p3}
\begin{array}{l}
 p\left( {{\bf{y}}_j \left| {{\bf{x}}_{j,m_j } ,{\bf{x}}_{1,n_1 } , \cdots ,{\bf{x}}_{j - 1,n_{j - 1} } ,{\bf{x}}_{j + 1,n_{j + 1} } , \cdots ,{\bf{x}}_{K,n_K } } \right.} \right) \\
  = \frac{1}{{\left( {\pi {\displaystyle \sigma} ^2 } \right)^{N_{r_j } } }}\exp \left( { - \frac{{\left\| {{\bf{y}}_j  - {\bf{H}}_{jj} {\bf{G}}_j {\bf{x}}_{j.m_j }  - \sum\limits_{i = 1,i \ne j}^K {{\bf{H}}_{ji} {\bf{G}}_i {\bf{x}}_{i.n_i } } } \right\|^2 }}{{{\displaystyle \sigma} ^2 }}} \right). \\
 \end{array}
\end{equation}

Plugging  (\ref{eqn:p1})--(\ref{eqn:p3}) into (\ref{eqn:H_y_3}) and (\ref{eqn:H_yx_4}) and define ${\mathbf{n}} = {\mathbf{y}}_j -
\sum\nolimits_{i = 1}^K {{\bf{H}}_{ji} {\bf{G}}_i {\bf{x}}_{i,m_i } }$, yields
\begin{equation}\label{eqn:H_y_4}
\begin{array}{l}
 H\left( {{\bf{y}}_j } \right) = \sum\limits_{i = 1}^K {\log_2 M_i} -
\frac{1}{{\prod\nolimits_{i = 1}^K {M_i } }}\sum\limits_{m_1  = 1}^{M_1}\sum\limits_{m_2  = 1}^{M_2}  \cdots \\
\qquad \qquad  \sum\limits_{m_K  = 1}^{M_K}{\int {p\left( {\bf{n}} \right)\log_2 H_{1,j} \left( {m_1 , m_2,\cdots ,m_K ,{\mathbf{n}}} \right)} }  d{\bf{n}}
\end{array}
\end{equation}
\begin{equation}\label{eqn:H_y_5}
\begin{array}{l}
\qquad \ \    = \sum\limits_{i = 1}^K {\log_2 M_i} - \frac{1}{{\prod\nolimits_{i = 1}^K {M_i } }}\sum\limits_{m_1  = 1}^{M_1} \sum\limits_{m_2  = 1}^{M_2} \cdots \\
\qquad \qquad \quad   \sum\limits_{m_K  = 1}^{M_K} {E_{\bf{n}} \left[ {\log_2 H_{1,j} \left( {m_1 , m_2 ,\cdots ,m_K ,{\mathbf{n}}} \right)} \right]}
\end{array}
\end{equation}

\begin{equation}\label{eqn:H_yx_5}
\begin{array}{l}
H\left( {{\bf{y}}_j \left| {{\bf{x}}_j } \right.} \right)  = \! \! \sum\limits_{i = 1, i \neq j}^K {\log_2 M_i} -
\frac{1}{{\prod\nolimits_{i = 1}^K {M_i } }}\sum\limits_{m_1  = 1}^{M_1} \sum\limits_{m_2  = 1}^{M_2}
\!  \cdots \!  \\  \sum\limits_{m_K  = 1}^{M_K}\int p\left( {\bf{n}} \right)\log_2 H_{2,j}
\left( {m_1 , \! \cdots \! ,m_{j-1},m_{j+1}, \! \cdots \! ,m_K, {\mathbf{n}} } \right)   d{\bf{n}}
\end{array}
\end{equation}
\begin{equation}\label{eqn:H_y_6}
\begin{array}{l}
\qquad \qquad \ = \sum\limits_{i = 1, i \neq j}^K {\log_2 M_i} - \frac{1}{{\prod\nolimits_{i = 1}^K {M_i } }}\sum\limits_{m_1  = 1}^{M_1} \sum\limits_{m_2  = 1}^{M_2} \cdots \\ {  \sum\limits_{m_K  = 1}^{M_K} {E_{\bf{n}} \left[ {\log_2 H_{2,j} \left( {m_1 , \cdots ,m_{j-1},m_{j+1},\cdots ,m_K, {\mathbf{n}} } \right)} \right]} }.
\end{array}
\end{equation}

Then, the proposition is proved by utilizing the definition of the mutual information \cite{Cover} $R_j = I({\mathbf{y}}_j,{\mathbf{x}}_j) = H(\mathbf{y}_j) - H\left( {{\bf{y}}_j \left| {{\bf{x}}_j } \right.} \right)$, along with the expressions in (\ref{eqn:H_y_5}) and (\ref{eqn:H_y_6}).


\section{Proof of Proposition \ref{prop:ach_rate_low}} \label{sec:proof_ach_rate_low}
First, we apply Jensen's inequality \cite{Boyd2004} to obtain an approximation of
(\ref{eqn:achievable_Rj}) as follows
\begin{eqnarray}\label{eqn:I_low_add_1}
 &&  R_{{{j}},{\rm{finite}}}  \approx \log M_j - \frac{1}{{\prod\nolimits_{i = 1}^K {M_i } }} \times \nonumber \\
 && \hspace{-0.8cm} \left[ \! f_{\rm{1}}\left( \! \!  E_{\bf{n}}{\exp\left( { \! - \! \frac{1}{{{\displaystyle \sigma} ^2 }}\left\| {\left( \! {\sum\limits_{i = 1}^K {{\bf{H}}_{ji} {\bf{G}}_i \left( {{\bf{x}}_{i,m_i }  \! \!- \! \!
   {\bf{x}}_{i,n_i } } \right)} } \! \right)\! \! + \!\! {\bf{n}}} \right\|^2 } \right)} \!  \right) \right. -  \nonumber \\
&& \hspace{-0.8cm}  \left.    f_{\rm{2}}\! \! \left( \!\! { E_{\bf{n}} {\exp \left( { \!\! - \frac{1}{{{\displaystyle \sigma} ^2 }}\left\|
  {\left( {\sum\limits_{i = 1,i \ne j}^K
   {{\bf{H}}_{ji}
   {\bf{G}}_i \left( {{\bf{x}}_{i,m_i }\!\! -\!\! {\bf{x}}_{i,n_i } } \right)} } \right) \!\! + \!\! {\bf{n}}} \right\|^2 } \right)} } \!\! \right) \!\! \right] \nonumber \\
\end{eqnarray}
where we define the operations $f_1(\cdot)$, $f_2(\cdot)$, $f_3(\cdot)$ and $f_4(\cdot)$ as
\begin{equation}\label{eqn:I_low_add_2}
f_1 (\cdot) \doteq  \sum\limits_{m_1  = 1}^{M_1 } \sum\limits_{m_2  = 1}^{M_2 } { \! \!\cdots \! \! \sum\limits_{m_K  = 1}^{M_K } {\log_2 } }  \sum\limits_{n_1  = 1}^{M_1 } \sum\limits_{n_2  = 1}^{M_2 } {\! \! \cdots \!\! \sum\limits_{n_K  = 1}^{M_K }}(\cdot)
\end{equation}
\begin{equation}\label{eqn:I_low_add_3}
f_2 (\cdot) \doteq \sum\limits_{m_1  = 1}^{M_1 } \sum\limits_{m_2  = 1}^{M_2 } \! \! \cdots \! \!\sum\limits_{m_K  = 1}^{M_K }
\! {\log_2 }  \! \sum\limits_{n_1  = 1}^{M_1 }  \! \! \cdots \! \! \sum\limits_{n_{j - 1}  = 1}^{M_{j - 1} } \sum\limits_{n_{j + 1}  = 1}^{M_{j + 1} }
 \! \! \cdots \! \! \sum\limits_{n_K  = 1}^{M_K }(\cdot)
\end{equation}
\begin{equation}\label{eqn:I_low_add_2}
f_3 (\cdot) \doteq  \sum\limits_{m_1  = 1}^{M_1 } \sum\limits_{m_2  = 1}^{M_2 } { \! \!\cdots \! \! \sum\limits_{m_K  = 1}^{M_K }  }  \sum\limits_{n_1  = 1}^{M_1 } \sum\limits_{n_2  = 1}^{M_2 } {\! \! \cdots \!\! \sum\limits_{n_K  = 1}^{M_K }}(\cdot)
\end{equation}
\begin{equation}\label{eqn:I_low_add_3}
f_4 (\cdot) \doteq \sum\limits_{m_1  = 1}^{M_1 } \sum\limits_{m_2  = 1}^{M_2 } \! \! \cdots \! \!\sum\limits_{m_K  = 1}^{M_K }
\!   \! \sum\limits_{n_1  = 1}^{M_1 }  \! \! \cdots \! \! \sum\limits_{n_{j - 1}  = 1}^{M_{j - 1} } \sum\limits_{n_{j + 1}  = 1}^{M_{j + 1} }
 \! \! \cdots \! \! \sum\limits_{n_K  = 1}^{M_K }(\cdot).
\end{equation}

Recalling the definition of ${\bf{n}}$ in (\ref{eqn:achievable_Rj}), the first expectation in  (\ref{eqn:I_low_add_1})
can be evaluated as
\[
\begin{array}{l}
 E_{\bf{n}} \exp\left( { - \frac{1}{{{\displaystyle \sigma} ^2 }}\left\| {\left( {\sum\limits_{i = 1}^K {{\bf{H}}_{ji} {\bf{G}}_i \left( {{\bf{x}}_{i,m_i }
   - {\bf{x}}_{i,n_i } } \right)} } \right) + {\bf{n}}} \right\|^2 } \right) \\
  = \int_{\bf{n}} \exp\left( { - \frac{1}{{{\displaystyle \sigma} ^2 }}\left\|
  {\left( {\sum\limits_{i = 1}^K {{\bf{H}}_{ji} {\bf{G}}_i \left( {{\bf{x}}_{i,m_i }  - {\bf{x}}_{i,n_i } } \right)} } \right) + {\bf{n}}}
   \right\|^2 } \right) \\
  \end{array}
 \]
\begin{equation}\label{eqn:I_low_01}
\begin{array}{l}
 \times  \exp \left( { - \frac{{\left\| {\bf{n}} \right\|^2 }}{{{\displaystyle \sigma} ^2 }}} \right)d{\bf{n}} \times \left( {\pi \sigma ^2 } \right)^{ - N_r }
  \end{array}      \\
\end{equation}
\begin{equation}\label{eqn:I_low_02}
\begin{array}{l}
  = 2^{ - N_r } \exp\left( { - \frac{1}{{2\sigma ^2 }}\left\| {\left( {\sum\limits_{i = 1}^K {{\bf{H}}_{ji} {\bf{G}}_i \left( {{\bf{x}}_{i,m_i }
   - {\bf{x}}_{i,n_i } } \right)} } \right)} \right\|^2 } \right) \\
  \qquad \qquad \qquad  \qquad \qquad  \qquad \times  \int_{\bf{n}} {I_1 \left( {\bf{n}} \right)d{\bf{n}}}  \\
    \end{array}
\end{equation}
where
\begin{eqnarray}\label{eqn:I_low_add}
&& \hspace{1.5cm} I_1 \left( {\bf{n}} \right) = \left( {\frac{{\pi \sigma ^2 }}{2}} \right)^{ - N_r }  \times  \nonumber \\
&& \hspace{-1cm} \exp\left( { - \frac{2}{{\sigma ^2 }}\left\|
{\left( {\frac{1}{2}\sum\limits_{i = 1}^K {{\bf{H}}_{ji} {\bf{G}}_i \left( {{\bf{x}}_{i,m_i }  - {\bf{x}}_{i,n_i } } \right)} } \right) + {\bf{n}}} \right\|^2 } \right)
\end{eqnarray}
denotes a probability density function of an independent and identically distributed Gaussian vector with mean
$-{\frac{1}{2}\sum\limits_{i = 1}^K {{\bf{H}}_{ji} {\bf{G}}_i \left( {{\bf{x}}_{i,m_i }  - {\bf{x}}_{i,n_i } } \right)} }$
and covariance matrix $\frac{\sigma ^2}{2} {\mathbf{I}}$. Therefore, we have
\begin{equation}\label{eqn:I_low_03}
\int_{\bf{n}} {I_1 \left( {\bf{n}} \right)d{\bf{n}}} = 1.
\end{equation}
Combining (\ref{eqn:I_low_01})--(\ref{eqn:I_low_03}) and applying similar approaches to the second expectation in (\ref{eqn:I_low_add_1}),
yields  $\tilde R_{j,{\rm{finite}}}$ as follows
\begin{eqnarray}\label{eqn:I_low_add_4}
 && \tilde R_{j,{\rm{finite}}}  =  \log M_j - \frac{1}{{\prod\nolimits_{i = 1}^K {M_i } }} \times \nonumber \\
  && \hspace{-0.5cm}\left[ \! f_{\rm{1}}\left( \! \!  {\exp\left( {  -  \frac{1}{{2 {\displaystyle \sigma} ^2 }}\left\| {\left( {\sum\limits_{i = 1}^K {{\bf{H}}_{ji} {\bf{G}}_i \left( {{\bf{x}}_{i,m_i }  \! \!- \! \!
   {\bf{x}}_{i,n_i } } \right)} } \! \right)} \right\|^2 } \right)} \!  \right) \right. -  \nonumber \\
 && \hspace{-0.5cm} \left.    f_{\rm{2}}\! \! \left( \!\! {  {\exp \left( { \!\! - \frac{1}{2 {{\displaystyle \sigma} ^2 }}\left\|
  {\left( {\sum\limits_{i = 1,i \ne j}^K
   {{\bf{H}}_{ji}
   {\bf{G}}_i \left( {{\bf{x}}_{i,m_i }\!\! -\!\! {\bf{x}}_{i,n_i } } \right)} } \right)  } \right\|^2 } \right)} } \!\! \right) \!\! \right]. \nonumber \\
\end{eqnarray}

It is noted that here we use the Jensen's inequality in
two terms in (\ref{eqn:achievable_Rj}) and then subtract them. When SNR is zero, the approximated rate $R_{j,\, \rm{finite}}$ in
(\ref{eqn:I_low_add_4}) is zero, which corresponds to the exact achievable rate in (\ref{eqn:achievable_Rj}). This indicates
that the approximation based on Jensen's inequality
is asymptotically accurate when SNR goes to zero. In addition,  since the ``bounding errors" for both terms are to be similar
and therefore subtracting these will have a canceling effect, which in turn
will yield a fairly accurate approximation.  This will be confirmed by numerical results where the obtained precoding design based
on (\ref{eqn:I_low_add_4}) performs nearly close to sum-rate achieved by Gaussian inputs in low SNR region,
which implies the obtained precoding design is actually near-optimal over all the possible solution sets.

In low SNR region where $\sigma^2 \rightarrow +\infty$, utilizing the Taylor expansion of the exponent function $\exp(x) = 1 + x + o(x)$,
  (\ref{eqn:I_low_add_4}) can be computed as
  \begin{eqnarray}\label{eqn:I_low_add_5}
 && \tilde R_{j,{\rm{finite}}}  =  \log M_j - \frac{1}{{\prod\nolimits_{i = 1}^K {M_i } }} \times \nonumber \\
  && \hspace{-0.5cm}\left[ \! f_{\rm{1}}\left( \! \!  1 - \frac{1}
   {2{\sigma ^2 }}{\left\| {\sum\limits_{i = 1}^K {{\bf{H}}_{ji} {\bf{G}}_i \left( {{\bf{x}}_{i,m_i } \! - \!{\bf{x}}_{i,n_i } }
  \! \right)\!} }  \right\|^2\! } + \!\! o\left( {\frac{1}{{\sigma ^2 }}}\! \right)
    \!  \right) \right. -  \nonumber \\
 && \hspace{-0.7cm} \left.    f_{\rm{2}}  { { \left( { \!\! 1 \! \!- \! \!\frac{1}{2 {{\displaystyle \sigma} ^2 }}\left\|
  {\left( {\sum\limits_{i = 1,i \ne j}^K
   {{\bf{H}}_{ji}
   {\bf{G}}_i \left( {{\bf{x}}_{i,m_i }\!\! -\!\! {\bf{x}}_{i,n_i } } \right)} } \right)  } \right\|^2 }\!\! + \!\! o\left( \!  {\frac{1}{{\sigma ^2 }}}  \right) \right)} }   \right]. \nonumber \\
\end{eqnarray}

Next, we exploit the Taylor expansion of function $\log_2(1 - x) = - \log_2 e \, x + o(x)$ to rewrite (\ref{eqn:I_low_add_5}) as
\begin{eqnarray}\label{eqn:I_low_add_5}
 && \hspace{-0.6cm} \tilde R_{j,{\rm{finite}}} \!  \! =   \! \!  \frac{{\log_2 e}}{2{\sigma ^2 \left( {\prod\nolimits_{i = 1}^K {M_i } } \right)^2 }} \!  \left[\!
\sum\limits_{i = 1}^K f_{\rm{3}} \left( \tr\left( {{\bf{H}}_{ji} {\bf{G}}_i \left( \!  {{\bf{x}}_{i,m_i }  \! \! -  \! \! {\bf{x}}_{i,n_i } } \! \right) }\right. \right. \right.  \nonumber \\
 && \times \left.{ \left( {{\bf{x}}_{i,m_i }  -  {\bf{x}}_{i,n_i } } \right)^H {\bf{G}}_i^H {\bf{H}}_{ji}^H } \right) - M_j  \sum\limits_{i = 1,i \ne j}^K  f_{\rm{4}} \left( \tr\left({{\bf{H}}_{ji} \! {\bf{G}}_i\! }\right. \right. \nonumber \\
 && \hspace{-0.5cm} \left. \! \times \left. {\left( {{\bf{x}}_{i,m_i } \! - \! {\bf{x}}_{i,n_i } } \right)
 \left( {{\bf{x}}_{i,m_i } \! - \! {\bf{x}}_{i,n_i } } \right)^H {\bf{G}}_i^H {\bf{H}}_{ji}^H } \right) \right] \! \! + \! \! o \left( \frac{1}{{\sigma ^2}} \right) .
\end{eqnarray}

For practical equiprobable symbols from symmetrical discrete constellation, we have
\begin{equation}\label{eqn:I_low_40}
\begin{array}{l}
\sum\limits_{m = 1}^{M_j } {{\bf{x}}_{j,m} }  {{\bf{x}}^H_{j,m} }  = M_j {\mathbf{I}},\quad
\sum\limits_{m = 1}^{M_j } {\sum\limits_{n = 1}^{M_j } { {{\bf{x}}_{j,m}   }  {{\bf{x}}^H_{j,n}}   } }  = 0, \\
\qquad \qquad \qquad \qquad \qquad \qquad  \qquad \qquad   j = 1,2,\cdots ,K.
\end{array}
\end{equation}

To this end,  (\ref{eqn:I_low_add_5}) can be further reduced to
\begin{eqnarray}\label{eqn:I_low_5}
 && \hspace{-0.6cm} \tilde R_{j,{\rm{finite}}} \! \! = \! \!\frac{{\log_2 e}}{{\sigma ^2 \left( {\prod\nolimits_{i = 1}^K {M_i } } \right)^2 }} \! \! \! \left( \! \left( \! {\prod\nolimits_{i = 1}^K {M_i } } \right)^2 \!\sum\limits_{i = 1}^K \tr\left( \!{{\bf{H}}_{ji} \!{\bf{G}}_i \!{\bf{G}}_i^H \!{\bf{H}}_{ji}^H } \!\right)\! \right. \nonumber \\
 && - \left. \left( {\prod\nolimits_{i = 1}^K {M_i } } \right)^2 \!\sum\limits_{i = 1,i \ne j}^K {\tr\left(\! {{\bf{H}}_{ji} \!{\bf{G}}_i\! {\bf{G}}_i^H \!{\bf{H}}_{ji}^H } \! \right)}   \right) \!+ \! o \left( \frac{1}{{\sigma ^2}} \right) \nonumber \\
 && \hspace{0.5cm}  = \frac{{\log_2 e}}{{\sigma ^2 }}\tr\left( {{\bf{H}}_{jj} {\bf{G}}_j {\bf{G}}_j^H {\bf{H}}_{jj}^H } \right) + o \left( \frac{1}{{\sigma ^2}} \right).
\end{eqnarray}

In (\ref{eqn:I_low_5}),  the achievable rate of each user is determined by its own channel gain and precoding matrix.
Also, it is straightforward to identify that the precoding design maximizing the first-order term in  (\ref{eqn:I_low_5}) is
to perform beamforming along the eigenvector corresponding to the largest eigenvalue of matrix
${\mathbf{H}}_{jj}^H{\mathbf{H}}_{jj}$, which completes the proof.

\section{ Proof of Corollary \ref{coro:low_snr} }\label{sec:Gaussian_low}
Here we prove that in low SNR region, the optimal precoding under Gaussian input assumption  conforms to the precoding
structure in Proposition \ref{prop:ach_rate_low}.   First, we rewrite achievable rate under Gaussian input assumption in (\ref{eqn:Rj}) as
\begin{equation}\label{eqn:Rj_new}
\begin{array}{l}
R_j \left( u \right) = \log_2 \det \left( {{\bf{I}} + u\sum\limits_{i = 1}^K {{\bf{H}}_{ji} {\bf{G}}_i
 {\bf{G}}_i^H {\bf{H}}_{ji}^H } } \right) \\
 \qquad \qquad  - \log_2 \det \left( {{\bf{I}} + u\sum\limits_{i = 1,i \ne j}^K
 {{\bf{H}}_{ji} {\bf{G}}_i {\bf{G}}_i^H {\bf{H}}_{ji}^H } } \right)
 \end{array}
\end{equation}
where $u = \frac{1}{{\sigma ^2 }}$.  In low SNR region, we expand $R_j \left( u \right)$ with respect to $u$ at $u = 0$ as
 \begin{equation}\label{eqn:Rj_low}
R_j \left( u \right) = \mathop {R_j }\limits^. \left( 0 \right)u + o\left( u \right).
\end{equation}
By exploiting \cite[Lemma 2]{Jafar2004TWC}, we have
 \begin{equation}\label{eqn:Rj_low_u}
\frac{d}{{du}}\log_2 \det \left( {{\bf{I}} + u{\bf{X}}} \right)\left| {_{u = 0} } \right. = \log_2 e \ \tr\left( {\bf{X}} \right).
\end{equation}

Combining (\ref{eqn:Rj_new})--(\ref{eqn:Rj_low_u}), yields
 \begin{equation}\label{eqn:Rj_low_2}
R_j  = \frac{{\log_2 e}}{{\sigma ^2 }} \tr\left( {{\bf{H}}_{jj} {\bf{G}}_j {\bf{G}}_j^H {\bf{H}}_{jj}^H } \right)  + o \left( \frac{1}{{\sigma ^2}} \right).
\end{equation}

We note that the precoding structure in Proposition \ref{prop:ach_rate_low} also maximizes the first-order term
in  (\ref{eqn:Rj_low_2}), which
completes our proof. Moreover,  the results in (\ref{eqn:I_low_5}) and (\ref{eqn:Rj_low_2}) implies
that (\ref{eqn:I_low_5}) actually
 provides  a first-order term upper bound for the achievable rate of each user with finite alphabet inputs in low SNR region.

\section{Proof of Proposition \ref{prop:ach_rate_high}} \label{sec:proof_ach_rate_high}
First, we consider $H_{1,j}(m_1,m_2,\cdots,m_K,{\bf{n}})$.  When $\sigma^2 \rightarrow 0$, we know if
${\sum\nolimits_{i = 1}^K {{\bf{H}}_{ji} {\bf{G}}_i \left( {{\bf{x}}_{i,m_i }  - {\bf{x}}_{i,n_i } } \right)} } \neq 0$,
it yields
 \begin{equation}\label{eqn:high_limit}
\lim_{\sigma^2 \rightarrow 0}{\exp \left( { - \frac{{\left\| {\left( {\sum\limits_{i = 1}^K {{\bf{H}}_{ji} {\bf{G}}_i \left( {{\bf{x}}_{i,m_i }  - {\bf{x}}_{i,n_i } } \right)} } \right) + {\bf{n}}} \right\|^2 }}{{\sigma ^2 }}} \right)} = 0.
\end{equation}
We consider the arrays $(m_1, m_2,\cdots,m_K) \neq (n_1, n_2,\cdots, n_K)$.
If $m_1 \neq n_1$,  we define
 \begin{equation}\label{eqn:high_c1}
c_{j,t}^{1,1} \left( {m_1 ,n_1 } \right) = \sqrt {\frac {P_1}{N_{t_1}} } a_{1,j,m_1 ,n_1 ,t}
\end{equation}
 \begin{equation}\label{eqn:high_c2}
 \begin{array}{l}
c_{j,t}^{1,2} \left( {m_2 ,m_3 , \cdots, m_K ,n_2 ,n_3 , \cdots, n_K } \right) \\
\qquad \qquad \qquad \qquad \qquad \qquad = \sum\limits_{i = 2}^K {\sqrt  \frac{\varepsilon _i P_i }{N_{t_i}} a_{i,j,m_i ,n_i ,t} }.
\end{array}
\end{equation}
Since $m_1 \neq n_1$, we can always find a $t^*  \in \left[ {1,N_{r_j} } \right]$, satisfies
\begin{equation}\label{ieq:high}
\left| {c_{j,t^* }^{1,1} \left( {m_1 ,n_1 } \right)} \right| \ge \sqrt {\frac {P_1}{N_{t_1}} } \omega _{1,\min }
\end{equation}
\begin{equation}\label{ieq:high_2}
\begin{array}{l}
\left| {c_{j,t^* }^{1,2} \left( {m_2 ,m_3 , \cdots ,m_K ,n_2 ,n_3 , \cdots ,n_K } \right)} \right| \\
\qquad \qquad \quad \mathop  \le \limits^{\left( a \right)} \sum\limits_{i = 2}^K {\sqrt  \frac{\varepsilon _i P_i }{N_{t_i}} \left| {a_{i,j,m_i ,n_i ,t^* } }\right|}
 \le \sum\limits_{i = 2}^K {\sqrt  \frac{\varepsilon _i P_i }{N_{t_i}} } \omega _{i,\max }
\end{array}
\end{equation}
where (a) is from the  Minkowski's inequality \cite{Saxe2002}.  Then, combining (\ref{ieq:high}), (\ref{ieq:high_2}), and the conditions in (\ref{Precoding_high_2}) ($i = 1$), we have
\begin{equation}\label{ieq:high_3}
\left| {c_{j,t^* }^{1,1} \left( {m_1 ,n_1 } \right)} \right| > \left| {c_{j,t^* }^{1,2} \left( {m_2 ,m_3 , \cdots ,m_K ,n_2 ,n_3 , \cdots ,n_K } \right)} \right|.
\end{equation}
(\ref{ieq:high_3}) implies that
\begin{equation}\label{ieq:high_4}
c_{j,t^* }^{1,1} \left( {m_1 ,n_1 } \right) \! + \! c_{j,t^* }^{1,2} \left( {m_2 ,m_3 ,\cdots, m_K ,n_2,n_3, \cdots, n_K } \right) \! \ne \! 0
\end{equation}
for arbitrary arrays $(m_1, m_2,\cdots,m_K) \neq (n_1, n_2,\cdots, n_K)$.

If $m_1 = n_1$ and  $m_2 \neq n_2$, we define
\begin{equation}\label{eq:high_c3}
c_{j,t}^{2,1} \left( {m_2 ,n_2 } \right) = \sqrt \frac{\varepsilon _2 P_2 }{N_{t_2}} a_{2,j,m_2 ,n_2 ,t }
\end{equation}
\begin{equation}\label{eq:high_c4}
\begin{array}{l}
c_{j,t }^{2,2} \left( {m_3 ,m_4 , \cdots ,m_K ,n_3 ,n_4 , \cdots ,n_K } \right) \\
\qquad \qquad \qquad \qquad \qquad  = \sum\limits_{i = 3}^K {\sqrt \frac{\varepsilon _i P_i }{N_{t_i}} a_{i,j,m_i ,n_i ,t } }.
\end{array}
\end{equation}
Following the similar approaches above, it obtains
\begin{equation}\label{ieq:high_5}
\left| {c_{j,t^* }^{2,1} \left( {m_2 ,n_2 } \right)} \right| > \left| {c_{j,t^* }^{2,2} \left( {m_3 ,m_4 , \cdots ,m_K ,n_3 ,n_4 , \cdots ,n_K } \right)} \right|
\end{equation}
\begin{equation}\label{ieq:high_5}
c_{j,t^* }^{2,1} \left( {m_2 ,n_2 } \right) + c_{j,t^* }^{2,2} \left( {m_3 ,m_4 , \cdots ,m_K ,n_3 ,n_4 , \cdots ,n_K } \right) \ne 0.
\end{equation}

This process continues until  $m_1  = n_1 ,m_2  = n_2 , \cdots ,m_{K - 1}  = n_{K - 1} ,m_K  \ne n_K $. We define
\begin{equation}\label{eq:high_c5}
c_{j,t }^{K,1} \left( {m_K ,n_K } \right) = \sqrt \frac{\varepsilon _K P_K }{N_{t_K}} a_{K,j,m_K ,n_K ,t}.
\end{equation}
Obviously we will find a $t^*  \in \left[ {1,N_{r_K} } \right]$ satisfying $c_{j,t^* }^{K,1} \neq 0$
due to the fact $m_K  \ne n_K$.

Thus, according to  the precoding design in (\ref{Precoding_high}), we
know that ${\sum\nolimits_{i = 1}^K {{\bf{H}}_{ji} {\bf{G}}_i \left( {{\bf{x}}_{i,m_i }  - {\bf{x}}_{i,n_i } } \right)} } \neq 0$ as long
as $(m_1, m_2,\cdots,m_K) \neq (n_1, n_2,\cdots, n_K)$.  This suggests
\begin{equation}\label{eq:high_lim_2}
\lim_{\sigma^2 \rightarrow 0} H_{1,j}(m_1,m_2,\cdots,m_K,{\bf{n}}) = \exp \left(-\frac{{\left\| {\bf{n}} \right\|^2 }}{{\sigma ^2 }}\right).
\end{equation}

Next, we consider $H_{2,j}(m_1,m_2,\cdots,m_K)$. Based on the precoding design in (\ref{Precoding_high}), if the equality ${\sum\nolimits_{i = 1, i \neq j}^K {{\bf{H}}_{ji} {\bf{G}}_i } } $ $ \left( {{\bf{x}}_{i,m_i } - {\bf{x}}_{i,n_i } } \right) = 0$
holds for any arrays $(m_1,\cdots,m_{j-1},m_{j+1}, \cdots, m_{K}) \neq
(n_1,\cdots,n_{j-1},$ $n_{j+1}, \cdots, n_{K})$, then by setting $m_j = n_j$, it yields
${\sum\nolimits_{i = 1}^K {{\bf{H}}_{ji} {\bf{G}}_i \left( {{\bf{x}}_{i,m_i }  - {\bf{x}}_{i,n_i } } \right)} } = 0 $, which contradicts with the conclusion above.
 As a consequence, we know that ${\sum\nolimits_{i = 1, i \neq j}^K {{\bf{H}}_{ji} {\bf{G}}_i \left( {{\bf{x}}_{i,m_i }  - {\bf{x}}_{i,n_i } } \right)} } \neq 0$ as long
as $(m_1,\cdots,m_{j-1},m_{j+1}, \cdots, m_{K}) \neq (n_1,\cdots,n_{j-1},$ $n_{j+1}, \cdots, n_{K})$. This implies
\begin{equation}\label{eq:high_lim_3}
\begin{array}{l}
\lim_{\sigma^2 \rightarrow 0} H_{2,j}(m_1,\cdots,m_{j-1},m_{j+1}, \cdots, m_{K}, {\bf{n}}) \\
\qquad \qquad \qquad \qquad \qquad \qquad \qquad  \qquad  = \exp \left(-\frac{{\left\| {\bf{n}} \right\|^2 }}{{\sigma ^2 }}\right).
\end{array}
\end{equation}

Combining (\ref{eqn:achievable_Rj}), (\ref{eq:high_lim_2}), and (\ref{eq:high_lim_3}), the achievable rate for the $j$-th user is given by
\begin{equation}\label{eq:high_lim_4}
\lim_{\sigma^2 \rightarrow 0} R_{j,\, \rm{finite}} = \log M_j, \ j = 1,2,\cdots, K.
\end{equation}
(\ref{eq:high_lim_4}) is the maximum rate that can be achieved for the $j$-th user
with respect to finite alphabet constraints, which completes
the proof.

It is noted that the optimal precoding design when $\sigma^2 \rightarrow 0$ is not unique.  For any precoders which fulfill the condition
${\sum\nolimits_{i = 1}^K {{\bf{H}}_{ji} {\bf{G}}_i \left( {{\bf{x}}_{i,m_i }  - {\bf{x}}_{i,n_i } } \right)} } \neq 0$ for arrays
$(m_1, m_2,\cdots,m_K) \neq (n_1, n_2,\cdots, n_K)$ can achieve the rate in (\ref{eq:high_lim_4}).

\section{Proof of Corollary \ref{coro:high_snr}}\label{sec:proof_high_snr}
Based on \cite[Theorem 3]{Gou2008}, we can transform the $K$ user MIMO interference channel in (\ref{eqn:model})
into a $K N_T$ user SIMO interference channel with $\eta$ antenna at each receiver. Let $\rho = K N_T \eta (K N_T - \eta - 1)$.
Then, recalling the interference alignment scheme in \cite[Appendix A]{Gou2008},
the length of desired signal transmitted over a $\nu_n = (\eta + 1) (n + 1)^\rho$ symbol extension for the $j$-th
user, $j = 1,2,\cdots,K$, is given by
\begin{equation}\label{eq:len_sym}
l_j  = \left\{ \begin{array}{lll}
 \eta \left( {n + 1} \right)^\rho , &j = 1,2, \cdots ,\eta  + 1 &\\
 \eta n^\rho  ,&j = \eta  + 2, \eta  + 3,\cdots ,K N_T& \\
 \end{array} \right.
\end{equation}
At the receiver, each user can decode its desired signal by zero forcing the aligned interference signals. Thus,
the sum-rate of all the receivers at high SNR is given by
\begin{equation}\label{eq:high_snr_rate}
\widetilde R_{{\rm{sum,  \, IA}}}^\infty (n)  = \eta \left( {n + 1} \right)^\rho  \sum\limits_{j = 1}^{\eta  + 1} {\log _2 q_j }  + \eta n^\rho  \sum\limits_{j = \eta  + 2}^{KN_T } {\log _2 q_j }
\end{equation}
where the elements of $q_j, j = 1,2,\cdots,K N_T$ are given by
\begin{equation}\label{eq:high_snr_q}
q_j  = Q_i , \quad \left( {i - 1} \right)N_T +1 \le j \le iN_T , \quad i = 1,2, \cdots, K.
\end{equation}
Then, the maximum average sum-rate per symbol can be computed as
\begin{equation}\label{eq:high_snr_avg_rate}
\overline{ R}_{{\rm{sum,IA}}}^\infty = \mathop {\sup}\limits_{n}
  \frac{1}{\nu_n}\widetilde R_{{\rm{sum,IA}}}^\infty (n).
\end{equation}
By plugging in the expressions of ${\nu_n}$ and $\widetilde R_{{\rm{sum,IA}}}^\infty (n)$ and taking the supremum over all $n$,
it yields
\begin{equation}\label{eq:high_snr_avg_rate}
\overline{R}_{{\rm{sum, \,IA}}}^\infty = \frac{\eta}{\eta + 1} \sum\limits_{j = 1}^{KN_T } \log_2 {q_j } =  \frac{\eta}{\eta + 1}  \sum\limits_{j = 1}^{K } \log_2 M_j.
\end{equation}

\section{Proof of Proposition \ref{prob:nec_cod}}\label{sec:proof_nec_cod}
The Lagrangian cost function for the precoding matrices in (\ref{eqn:WSR_1})--(\ref{eqn:WSR_2}) is given by
\begin{equation}\label{eqn:cost_function}
\begin{array}{l}
f\left( {{\bf{G}}, {\boldsymbol{\kappa}}} \right) =  - R_{{\rm{wsum,\,finite}}} \left( {{\bf{G}}_1 ,{\bf{G}}_2,
\cdots ,{\bf{G}}_K } \right) \\
\qquad \qquad \qquad \qquad  \qquad \qquad + \sum\limits_{j = 1}^K {\kappa _j \left[ {\tr\left( {{\bf{G}}_j {\bf{G}}_j^H } \right) - P_j } \right]}
\end{array}
\end{equation}
in which $\kappa _j \geq 0, j = 1,2,\cdots, K$. Following the similar approaches in \cite{Lozano2006TIT},
we define the complex gradient operator as $\nabla _{{\bf{G}}_j} f = \frac{{\partial f}}{{\partial {\bf{G}}_j^* }}$. The $\left( {i,j} \right)^{{\rm{th}}}$
element of matrix with the gradient is defined as $\left\{ {\nabla _{{\bf{G}}_j} f} \right\}_{i,j}  = \nabla _{\left\{ {\bf{G}}_j \right\}_{i,j} }
f = \frac{{\partial f}}{{\partial \left\{ {{\bf{G}}_j^* } \right\}_{i,j} }}$.  Then, the KKT conditions in \cite{Boyd2004} are as follows
\begin{equation}\label{eqn:KKT_1}
\begin{array}{l}
\nabla _{{\bf{G}}_j } f\left( {{\bf{G}},{\boldsymbol{\kappa}} } \right)=
  - \nabla _{{\bf{G}}_j } R_{{\rm{wsum,finite}}} \left( {{\bf{G}}_1 , {\bf{G}}_2, \cdots ,{\bf{G}}_K } \right)  \\
  \qquad \qquad \qquad \qquad \qquad \qquad \qquad \qquad   + \kappa _j {\bf{G}}_j = 0
\end{array}
\end{equation}
\begin{equation}\label{eqn:KKT_2}
\kappa _j \left[ {\tr\left( {{\bf{G}}_j {\bf{G}}_j^H } \right) - P_j } \right] = 0
\end{equation}
\begin{equation}\label{eqn:KKT_3}
\tr \left( {{\bf{G}}_j {\bf{G}}_j^H } \right) - P_j  \le 0
\end{equation}
\begin{equation}\label{eqn:KKT_4}
\kappa _j  \geq 0
\end{equation}
for all $j = 1,2,\cdots, K$, where (\ref{eqn:KKT_1}) is obtained through the complex matrix differentiation results  in \cite{Hjorungnes2011}.

Next, we consider the calculation of  $\nabla _{{\bf{G}}_j } R_{{\rm{wsum,\, finite}}} \left( {\bf{G}}_1, \right.$  $ \left. {\bf{G}}_2, {\cdots ,{\bf{G}}_K } \right)$.
For $ R_{j} \left( {\bf{G}}_1 ,{\bf{G}}_2, \cdots, \right. $ $\left. { {\bf{G}}_K } \right)$ in (\ref{eqn:WSR_1}), we know that
$H_{2,j} \left( {m_1 , \cdots ,m_{j - 1}
,m_{j + 1} }\right. $ $\left.{, \cdots ,m_K },{\bf{n}} \right)$ is independent of ${\bf{G}}_j$ from (\ref{eqn:achievable_H_2}).
For $R_{i} \left( {{\bf{G}}_1 ,{\bf{G}}_2, \cdots }\right.$ $\left.{,{\bf{G}}_K } \right), i \neq j$ in (\ref{eqn:WSR_1}),
both $H_{1,j} \left( {m_1 , m_2 , } \right.$ $ \left. { \cdots ,m_K ,{\bf{n}}} \right)$ and
$H_{2,j} \left( {m_1 , \cdots ,m_{j - 1} ,m_{j + 1} , \cdots ,m_K,{\bf{n}}} \right)$ are functions of ${\bf{G}}_j$.
Then,  by exploiting
the matrix derivative technology  in \cite{Petersen}, along with some simplifications, we can obtain the derivative expressions
$\nabla _{{\bf{G}}_j } R_{{\rm{wsum,\, finite}}} \left( {\bf{G}}_1, \right.\left. {\bf{G}}_2, {\cdots ,{\bf{G}}_K } \right)$,
given by the right-hand-side of (\ref{eqn:nec_cod_1}).



\end{document}